\documentclass[reprint,aps,pre]{revtex4-2}
\usepackage{graphicx}
\usepackage{amssymb}
\usepackage{amsmath}
\usepackage{gensymb}
\usepackage{xcolor}
\usepackage{hyperref}
\usepackage[utf8]{inputenc}
\usepackage{comment}
\usepackage{lipsum}
\usepackage[normalem]{ulem}
\usepackage{orcidlink}
\usepackage{subfigure}
\usepackage[section]{placeins}
\usepackage[margin=1.5cm]{geometry}
\begin{document}
\title{Emergence of correlation-driven altermagnetism in Hubbard model on geometrically
frustrated lattice-clusters}
\author{Md Fahad Equbal \orcidlink{0009-0004-0054-9068}}
 \email{mfequbal33@gmail.com (Corresponding Author)}
 \affiliation{Department of Physics, Jamia Millia Islamia (Central University), New Delhi $110025$, India}
\author{M. A. H. Ahsan \orcidlink{0000-0002-9870-2769}}
 \email{mahsan@jmi.ac.in}
\affiliation{Department of Physics, Jamia Millia Islamia (Central University), New Delhi $110025$, India}
\begin{abstract}
We investigate the emergence of correlation-driven altermagnetism in simple and extended Hubbard model on geometrically frustrated lattice-clusters using exact diagonalization. By systematically tuning the degree of geometric frustration across different cluster geometries---including the fully frustrated $3\times3$ torus, the partially frustrated $2\times3$ cylinder and the unfrustrated $2\times4$ cylindrical lattice---we isolate the necessary conditions for compensated anisotropic spin order. At half-filling on $3\times3$ lattice, the average altermagnetic spin-response $\langle \Delta_{\mathrm{spin}} \rangle$ remains small despite robust local moment formation, indicating that localized moments alone are insufficient to break directional symmetry. In contrast, the introduction of a single mobile charge carrier (hole or electron) induces a finite $\langle \Delta_{\mathrm{spin}} \rangle$ that increases monotonically with on-site interaction $U$. This altermagnetic phase is characterized by a $d$-wave-like alternating sign structure in real-space correlations, a broad momentum-space distribution in the altermagnetic structure factor $S_{\mathrm{alm}}(\mathbf{q})$ and a distinct particle-hole asymmetry with significantly enhanced response for one-electron-doped system. We demonstrate that these correlations remain zero on the unfrustrated $2\times4$ lattice, establishing geometric frustration as a prerequisite for altermagnetism. The inclusion of nearest-neighbor Coulomb repulsion $V$ weakens the altermagnetic correlations in $3\times3$ lattice through enhanced electronic localization but facilitates the onset of altermagnetic order in $2\times3$ lattice beyond a critical threshold. Finally, we show that the anisotropy parameter $\mathcal{A}$ is non-zero only for degenerate ground states, revealing that macroscopic symmetry breaking on finite clusters requires a confluence of geometric frustration and ground-state degeneracy. Our results establish a purely interaction-driven mechanism for altermagnetism, arising from the interplay of geometric frustration, carrier itinerancy and the competition between local and nonlocal Coulomb repulsions.

{\bf Keywords}: altermagnetism; electronic correlations; geometric frustration; Hubbard model; exact diagonalization
\end{abstract}
\date{\today}
\maketitle
\section{Introduction}
\label{intro} 
Strongly correlated electron systems remain at the forefront of condensed matter physics due to the rich interplay between charge, spin and lattice degrees of freedom \cite{Edagotto1994, Edagotto2005}, which gives rise to emergent quantum phases such as Mott insulators \cite{Imada1998, Hpark2008, Maria2024, Mfequbal2026}, unconventional superconductors \cite{Kirtley2000, Efradkin2015, Keimer2015, Fahad2025s}, quantum spin liquids \cite{Savary2017, Broholm2020, Jiang2021} and exotic magnetic states \cite{Ssalunke2007, Mahfoozul2016, Drewes2017, Werner2020, Piotr2021}. The Hubbard model \cite{Hubbard1963, Gutzwiller1963, Kanamori1963} and its extensions \cite{Micnas1990} provide a minimal microscopic framework for capturing these phenomena, interpolating between itinerant and localized regimes through the competition between kinetic energy, on-site Coulomb repulsion and nonlocal interactions. In the strong-coupling limit, the model maps onto effective Heisenberg-type spin Hamiltonian \cite{Arai1962, Tusch1996}, while at intermediate coupling it retains simultaneous charge and spin fluctuations making it particularly sensitive to carrier doping (hole or electron) and geometric frustration \cite{Luca2020, Huang2024}.

Recent years have witnessed growing interest in altermagnetism, a class of collinear magnetic states that exhibit momentum-dependent spin splitting despite vanishing net magnetization and preserved combined symmetries \cite{Hayami2019, Hayami2020J, Hayami2020O} distinct from conventional ferromagnets and antiferromagnets \cite{smejkalprx2022, smejkal2022, Igor2022, Jungwirth2025}. Unlike conventional antiferromagnets, where spin-degenerate bands are protected by combined time-reversal and translation or inversion symmetries \cite{Zhou2025}, altermagnets display momentum-dependent spin splitting analogous to $d$-, $g$-, or $i$-wave symmetries \cite{Sim2025, Karetta2025, Annica2025}. This unique combination of properties positions altermagnets as promising platforms for next-generation spintronic devices, including terahertz oscillators, spin-torque elements, and topological quantum applications \cite{Zexin2022, Bose2022, Thomas2023, Ling2024, Lu2025, Song2025}. The identification of altermagnetic signatures in correlated systems has opened new perspectives on how electron-electron interactions can give rise to symmetry-selective spin responses even in centrosymmetric lattices \cite{Linding2020, Aldo2024, Giuli2025, Cheong2025}.

From a microscopic viewpoint, the emergence of altermagnetic correlations is intimately connected to the interplay between local moment formation and the spatial organization of spin correlations \cite{Leeb2024, Scheong2024}. Geometric frustration plays a central role in suppressing conventional long-range magnetic order and promoting unconventional correlated states \cite{Moro2023}. On frustrated lattices, competing exchange pathways prevent simple bipartite N\'eel ordering, often stabilizing highly degenerate manifolds \cite{Zjing2022, Krivnov2025}, spin-liquid-like correlations \cite{Aramirez2022, Binbin2022} or spatially modulated spin textures \cite{Yliu2024}. 

\begin{figure}[h]
\centering
\includegraphics[scale=0.65]{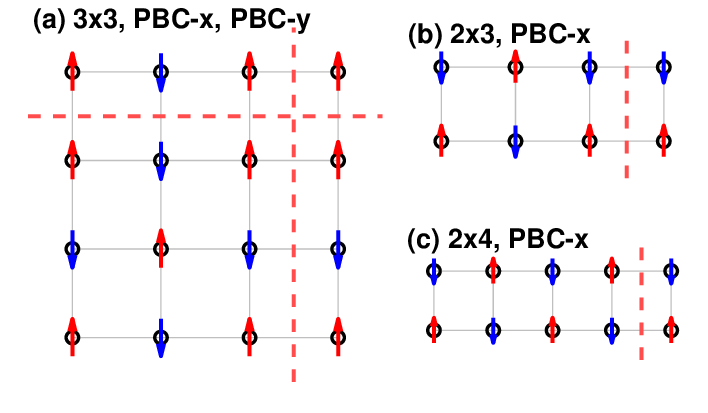}
\caption{Schematic illustration of the lattice geometries considered: (a) fully frustrated $3\times 3$ square lattice with periodic boundary conditions in both $x$- and $y$-directions (torus geometry), (b) partially frustrated $2\times 3$ lattice with periodic boundary conditions along the $x$-direction (cylindrical geometry), and (c) unfrustrated $2\times 4$ lattice with periodic boundary conditions along the $x$-direction (cylindrical geometry). Red dashed lines indicate periodic connections.}
\label{fig:Fig1}
\end{figure} 

In this work, we investigate the simple and the extended Hubbard model using exact diagonalization (ED) \cite{Fahad2025ps, Fahad2026ps} on finite-size lattices with distinct geometries and boundary conditions that together tune the degree of geometric frustration. Specifically, we consider (i) a fully frustrated $3\times 3$ square lattice with periodic boundary conditions (PBC) in both spatial directions (torus geometry), (ii) a partially frustrated $2\times 3$ lattice with PBC along one direction (cylindrical geometry), and (iii) an unfrustrated $2\times 4$ lattice with cylindrical boundary conditions, as shown in Fig. \ref{fig:Fig1}. While the $3\times 3$ lattice is studied at half-filling ($N=9$) and in the presence of single-carrier (hole or electron) doping ($N=8,10$), the $2\times 3$ and $2\times 4$ lattices are analyzed at half-filling to isolate the role of geometric frustration independent of carrier dynamics. This comparative framework enables us to disentangle the interplay between geometric frustration, carrier itinerancy and interaction effects across intermediate ($U=4$) and strong-coupling ($U=10$) regimes.

The choice of these lattice geometries provides a controlled framework for investigating the emergence of frustration-driven altermagnetic correlations. The $3\times 3$ toroidal lattice represents a minimal fully frustrated system where PBC in both spatial directions and odd number of lattice sites suppress conventional N\'eel-type spin ordering and promote a highly degenerate manifold of competing spin configurations. The $2\times 3$ cylindrical lattice retains frustration only along one spatial direction reducing this degeneracy whereas the $2\times 4$ lattice is bipartite and free from geometric frustration favoring conventional antiferromagnetic correlations. Together, these geometries span the full range of frustration---from maximal (odd-by-odd torus) to intermediate (cylindrical with one frustrated direction) to absent (even-by-even lattice with cylindrical boundary conditions)---while remaining computationally tractable within ED. This progression allows us to examine whether the observed altermagnetic correlations arise intrinsically from electronic interactions or instead depend on the frustration-induced suppression of conventional long-range antiferromagnetic order.

The remainder of this manuscript is organized as follows. In Sec. \ref{modmet}, we introduce the model Hamiltonian and the formalism used to study correlation-driven altermagnetism. Section \ref{resdis} presents our results, beginning with the simple Hubbard model ($V=0$) to establish the role of doping and geometric frustration followed by an analysis of the extended Hubbard model at intermediate ($U=4$) and strong ($U=10$) coupling. Finally, Sec. \ref{summary} summarizes our findings.

\section{Model and method}
\label{modmet}
We study the emergence of altermagnetic correlations in the simple and the extended Hubbard model on finite-size two-dimensional (2D) lattices using ED. The one-band extended Hubbard Hamiltonian is given by \cite{Micnas1990} 
\begin{equation}
H=-t\sum_{\langle ij\rangle\sigma}(c_{i\sigma}^\dagger c_{j\sigma}+h.c.)+U\sum_i n_{i\uparrow}n_{i\downarrow}+V\sum_{\langle ij \rangle}n_{i} n_{j},
\label{hamil}
\end{equation}
where $i$ and $j$ label lattice sites and $\langle ij\rangle$ denotes nearest-neighbor (NN) pairs. The operator $c_{i\sigma}^\dagger (c_{i\sigma})$ creates(annihilates) an electron with spin $\sigma\in\lbrace \uparrow, \downarrow\rbrace$ on site $\textit{i}$, $n_{i\sigma}=c_{i\sigma}^\dagger c_{i\sigma}$ is the fermion number operator, $n_i=n_{i\uparrow}+n_{i\downarrow}$ and $h.c.$ means Hermitian conjugate. The parameters $t$, $U$ and $V$ denote the NN hopping amplitude, on-site Coulomb interaction and NN Coulomb interaction, respectively. In the limit $V=0 $, Eq. \ref{hamil} reduces to the simple Hubbard model. In the non-interacting limit ($U=V=0$), the dispersion relation is $\epsilon_{\mathbf{k}}=-2t(\cos k_x+\cos k_y)$ with bandwidth $W=8t$ in the thermodynamic limit.
 
We perform calculations on the lattice geometries introduced in Sec. \ref{fig:Fig1}. For the $3\times 3$ lattice, we consider half-filling ($N=9$) as well as single-carrier doping with one hole ($N=8$) and one electron ($N=10$). For the $2\times 3$ and $2\times 4$ lattices, calculations are restricted to half-filling. In all cases, we focus on the lowest total-spin sector, namely $S=\frac{1}{2}$ for $N=9$ and $S=0$ for even fillings, thereby restricting the analysis to magnetically compensated states. 

The Hamiltonian is diagonalized exactly in spin-adapted basis \cite{Mahsan1994, Sarma1996}, which fully resolves the total spin $S$ by exploiting the $SU(2)$ symmetry of the model, $[H,S^2]=[H,S^z]=0$. This approach significantly reduces the Hilbert-space dimensionality and enables direct access to spin-resolved correlation functions. For each set of parameters $(U,V)$, we compute the ground state and low-lying excited states, from which equal-time correlation functions are obtained.

To characterize altermagnetic correlations, we introduce site-resolved correlation matrix constructed from equal-time spin-spin correlations on nearest-neighbor bonds
\begin{equation}
\Omega_{ij}
=
\frac{\Omega_0}{2}
\langle \mathbf{S}_i \cdot \mathbf{S}_j \rangle
\left(
\delta_{j,i\pm a\hat{\mathbf e}_x}
-
\delta_{j,i\pm a\hat{\mathbf e}_y}
\right),
\label{almordmat}
\end{equation}
where the Kronecker delta symbol is defined as

\begin{equation}
\delta_{j,i\pm a\hat{\mathbf e}_{\mu}}
=
\begin{cases}
1, & \mathbf r_j=\mathbf r_i\pm a\hat{\mathbf e}_{\mu},\\[4pt]
0, & \text{otherwise},
\end{cases}
\qquad \mu=x,y.
\label{bondkronecker}
\end{equation}
Here $\mathbf{S}_i$ is the spin operator at site $i$, $\Omega_0$ is a normalization constant, $\mathbf{r}_i$ is the position of site $i$, $a$ is the lattice spacing and $\hat{\mathbf{e}}_x$ ($\hat{\mathbf{e}}_y$) is the unit vector along the $x$ ($y$) direction. The indices $i$ and $j$ label lattice sites such that $\Omega_{ij}$ represents the correlation matrix element associated with the site pair $(i,j)$; no Einstein summation convention is assumed. The Kronecker delta symbols restrict the correlations to nearest-neighbor bonds and assign opposite signs to bonds oriented along the $x$ and $y$ directions, thereby encoding the $d$-wave-like sign structure characteristic of altermagnetic order.

The corresponding momentum-space signature is captured by the altermagnetic structure factor
\begin{equation}
S_{\mathrm{alm}}(\mathbf{q}) = \frac{1}{M} \sum_{i,j=1}^M \Omega_{ij} e^{i \mathbf{q}\cdot(\mathbf{r}_i - \mathbf{r}_j)},
\label{almsf}
\end{equation}
where $M$ is the total number of lattice sites and $\mathbf{q}$ denotes the discrete wavevectors permitted by finite lattice.

In finite-size ED framework, where spin-resolved band dispersions are not well defined, altermagnetic tendencies are characterized through real-space correlation measures. We thus define site-resolved spin-response measure
\begin{equation}
\Delta_{i}^{\mathrm{spin}}=\frac{1}{N_b(i)}\sum_{j=1}^M \Omega_{ij},
\label{ssi}
\end{equation}
where $N_{b}(i)$ is the number of nearest-neighbor bonds connected to site $i$. The corresponding average spin-response measure is 
\begin{equation}
\langle\Delta_{\mathrm{spin}}\rangle=\frac{1}{M}\sum_{i=1}^M \Delta_{i}^{\mathrm{spin}},
\label{ssg}
\end{equation}
which serves as a diagnostic of altermagnetic correlations by enhancing compensated, anisotropic spin-correlation patterns. 

To distinguish correlation effects from changes in local spin magnitude, we compute the average local moment
\begin{equation}
\bar{m}=\frac{4}{N}\sum_i\langle(n_{i\uparrow}-n_{i\downarrow})^2\rangle
\label{alclm}
\end{equation}
which quantifies the degree of electron localization.

Finally, we quantify directional anisotropy via the parameter
\begin{equation}
\mathcal{A}=\frac{1}{M}\sum_i\left|\Delta_i^{x}-\Delta_i^{y}\right|,
\label{anisotropy}
\end{equation} 
where $\Delta_i^{x(y)}$ is obtained by restricting the sum in Eq.~\ref{ssi} to bonds along the $x(y)$ direction.
\section{Results and discussion}
\label{resdis}
This section presents the results for altermagnetic correlations and their evolution across different interaction regimes. The simple Hubbard model ($V=0$) is investigated first, focusing on the interplay between on-site interaction, geometric frustration and carrier doping. Subsequently, the role of the nearest-neighbor (NN) Coulomb interaction $V$ in modifying these correlations is examined within the extended Hubbard model. Throughout this work, energies are measured in units of the hopping amplitude $t=1$.
\subsection{Simple Hubbard Model ($V=0$)}
\label{shm}
\begin{figure}[h]
\centering
\includegraphics[scale=0.62]{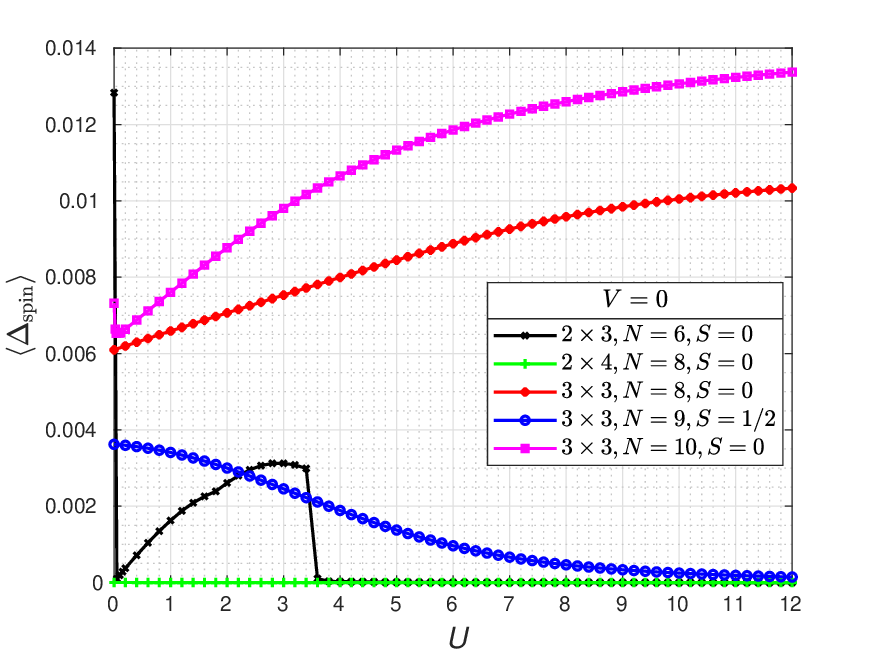}
\caption{Average altermagnetic spin-response $\langle \Delta_{\mathrm{spin}} \rangle$ as a function of on-site interaction $U$ for the simple Hubbard model ($V=0$). Results are shown for the fully frustrated $3\times3$ lattice at half-filling ($N=9$) and with one-hole ($N=8$) and one-electron ($N=10$) doping, the partially frustrated $2\times3$ lattice at half-filling ($N=6$) and the unfrustrated $2\times4$ lattice at half-filling ($N=8$).}
\label{fig:Fig2}
\end{figure}

We begin by examining the evolution of the average altermagnetic spin-response $\langle \Delta_{\mathrm{spin}} \rangle$ in the simple Hubbard model ($V=0$). The results are summarized in Fig. \ref{fig:Fig2}. For half-filled systems, the behavior of the average altermagnetic spin-response $\langle \Delta_{\mathrm{spin}} \rangle$ is strongly controlled by lattice geometry. On the unfrustrated $2\times4$ lattice ($N=8$, $S=0$), $\langle \Delta_{\mathrm{spin}} \rangle$ remains zero for all values of $U$, reflecting bipartite symmetry of the lattice, where conventional N\'eel-type antiferromagnetic correlations dominate. In this case, the staggered structure encoded in the altermagnetic spin correlation operator $\Omega_{ij}$ [Eq. (\ref{almordmat})] leads to a complete cancellation of the average altermagnetic spin-response, indicating the absence of directional anisotropy in the spin correlations. Introducing geometric frustration qualitatively modifies this picture. On the fully frustrated $3\times3$ lattice at half-filling ($N=9$, $S=1/2$), $\langle \Delta_{\mathrm{spin}} \rangle$ remains finite at weak coupling, signaling the emergence of frustration-induced anisotropic spin correlations. As $U$ increases, however, $\langle \Delta_{\mathrm{spin}} \rangle$ is progressively suppressed. In the strong-coupling regime, the system develops robust local moments but explores a highly degenerate spin manifold. While geometric frustration prevents simple N\'eel order, the resulting correlations lack a coherent directional structure, leading to a reduction of the altermagnetic spin response. The partially frustrated $2\times3$ lattice at half-filling ($N=6$, $S=0$) exhibits a distinct non-monotonic dependence on $U$. The average altermagnetic spin-response $\langle \Delta_{\mathrm{spin}} \rangle$ increases at weak coupling, reaches a maximum near $U \approx 3.4$, and then drops sharply at $U_c \approx 3.6$. This behavior signals a crossover from an itinerant regime with enhanced anisotropic spin correlations to a localized phase in which these correlations are suppressed.

A qualitatively different trend is observed upon doping the fully frustrated $3\times3$ lattice. For both one-hole ($N=8$) and one-electron ($N=10$) doped systems, $\langle \Delta_{\mathrm{spin}} \rangle$ increases monotonically with $U$ and saturates at strong coupling, indicating that mobile carriers stabilize anisotropic spin correlations. The average altermagnetic spin-response $\langle \Delta_{\mathrm{spin}} \rangle$ is consistently large in the electron-doped system, revealing particle-hole asymmetry associated with the underlying band structure of the geometrically frustrated lattice. 
Overall, these results show that a finite altermagnetic spin-response arises from the combined effects of geometric frustration and carrier itinerancy. In their absence---either in an unfrustrated geometry or in the localized half-filled regime---the altermagnetic response is suppressed.

\begin{figure}[h]
\centering
\includegraphics[scale=0.62]{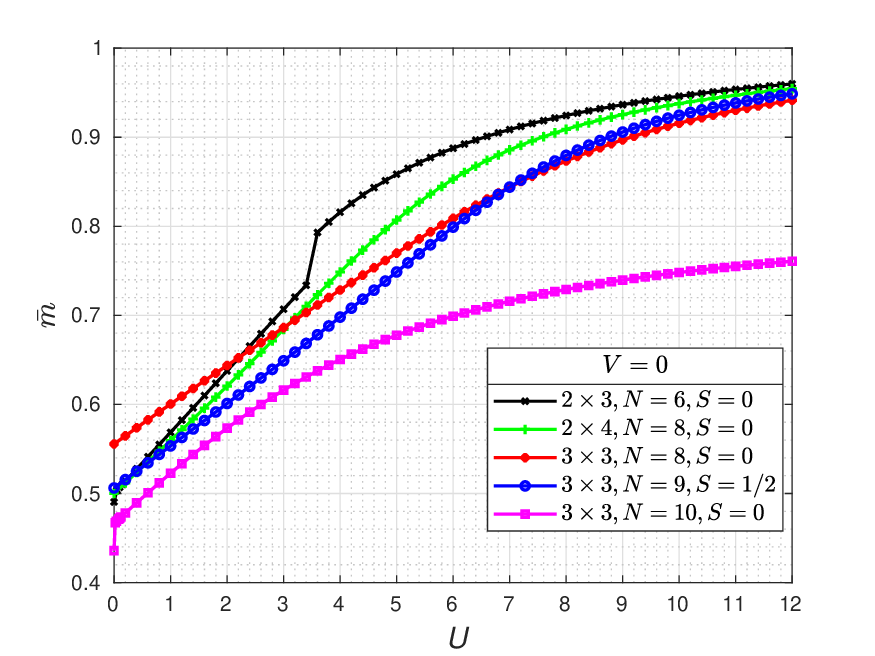}
\caption{Average local moment $\bar{m}$ as a function of the on-site interaction $U$ for the simple Hubbard model ($V=0$). Results are shown for the fully frustrated $3\times3$ lattice at half-filling ($N=9$) and with one-hole ($N=8$) and one-electron ($N=10$) doping, the partially frustrated $2\times3$ lattice at half-filling ($N=6$), and the unfrustrated $2\times4$ lattice at half-filling ($N=8$).}
\label{fig:Fig3}
\end{figure}

To elucidate the microscopic origin of the interaction-driven altermagnetic response, we examine the average local moment $\bar{m}$ as a function of $U$, shown in Fig. \ref{fig:Fig3}. Across all geometries, $\bar{m}$ increases monotonically with $U$, reflecting the suppression of double occupancy by the on-site Coulomb repulsion. The detailed evolution, however, depends sensitively on lattice geometry and electron filling.

For half-filled systems, geometric frustration modulates the growth of $\bar{m}$. On the unfrustrated $2\times4$ lattice ($N=8$, $S=0$), $\bar{m}$ increases smoothly and saturates rapidly, consistent with conventional Mott localization on a bipartite lattice. On the fully frustrated $3\times3$ lattice ($N=9$, $S=1/2$), the growth of $\bar{m}$ is more gradual, indicating that geometric frustration sustains quantum fluctuations and delays complete localization. In the partially frustrated $2\times3$ lattice ($N=6$, $S=0$), $\bar{m}$ exhibits a sharp increase near $U_c \approx 3.6$. This non-analytic feature coincides with the abrupt drop in $\langle\Delta_{\mathrm{spin}}\rangle$ observed in Fig. \ref{fig:Fig2}, indicating a crossover from an itinerant regime with anisotropic spin correlations to a localized phase.

Turning to the doped cases of the fully frustrated $3\times3$ lattice, both one-hole-doped ($N=8$) and one-electron-doped ($N=10$) systems display a slower increase of $\bar{m}$ with $U$ compared to the half-filled ($N=9$) system, reflecting persistent charge fluctuations. The one-electron-doped system consistently exhibits the smallest local moment across the entire interaction range, indicating the highest degree of itinerancy. This correlates directly with the enhanced altermagnetic response observed in Fig. \ref{fig:Fig2}, demonstrating that residual itinerancy is essential for sustaining anisotropic spin correlations at strong coupling.

In the large-$U$ limit, $\bar{m}$ approaches saturation for the half-filled systems ($N=6, 8, 9$), consistent with the formation of localized moments. This behavior accounts for the suppression of $\langle\Delta_{\mathrm{spin}}\rangle$ in this regime, as increasing localization reduces the spatial coherence of directional spin correlations. Collectively, these results show that the altermagnetic response is maximized in an intermediate regime where local moments are established but electronic motion remains appreciable, whereas strong localization suppresses the directional anisotropy required to sustain altermagnetic correlations.

\begin{figure}[h]
\centering
\includegraphics[scale=0.34]{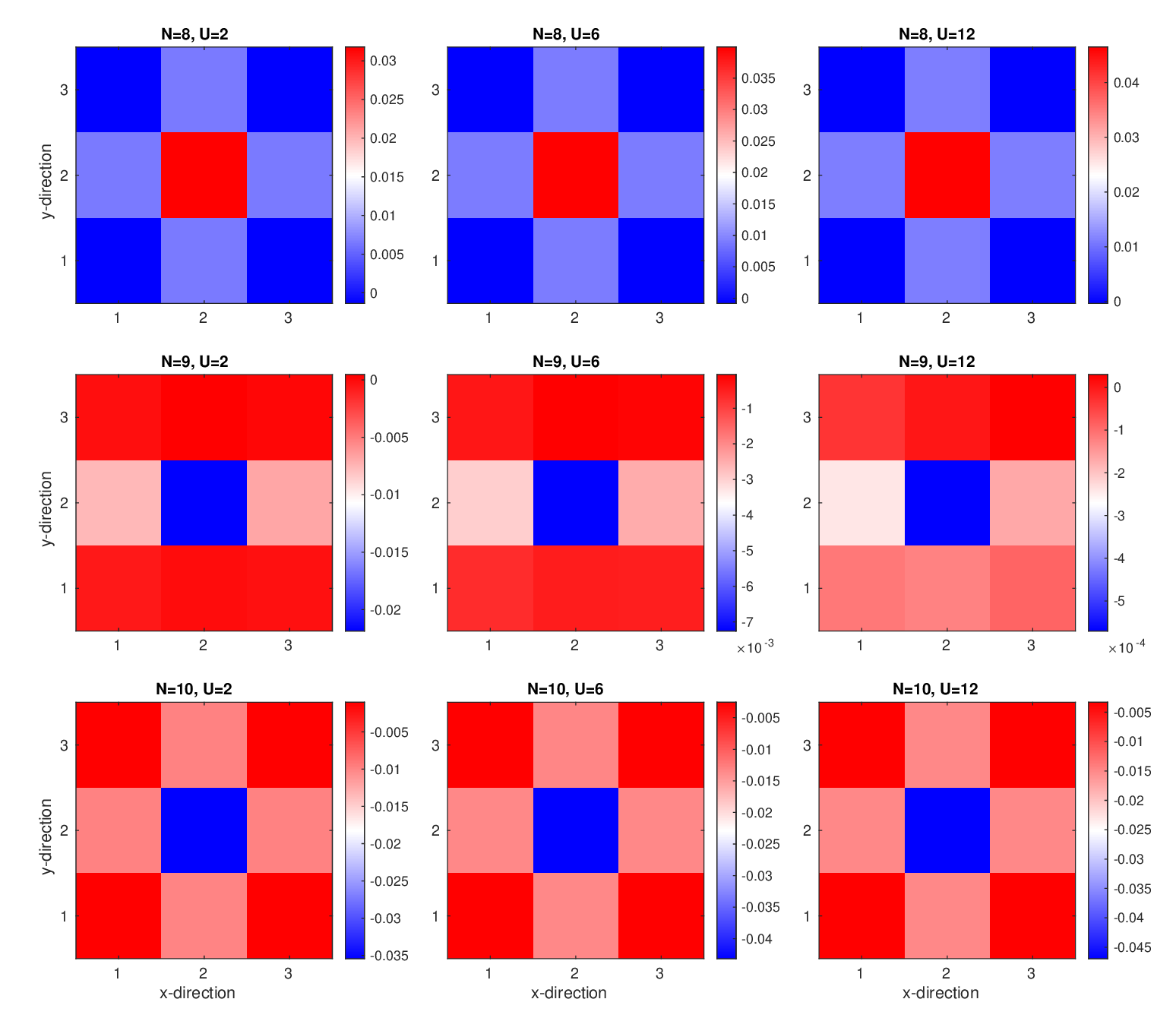}
\caption{Real-space heatmaps of the altermagnetic correlation matrix elements $\Omega_{5j}$ relative to the central site ($i=5$) for the fully frustrated $3\times3$ lattice. Rows correspond to one-hole doping ($N=8$, top), half-filling ($N=9$, middle) and one-electron doping ($N=10$, bottom), while columns show increasing interaction strength $U=2,6$ and 12.}
\label{fig:Fig4}
\end{figure}

To visualize the spatial organization of altermagnetic correlations, Fig. \ref{fig:Fig4} presents heatmaps of the correlation matrix elements $\Omega_{5j}$ with respect to the central site ($i=5$) for the fully frustrated $3\times3$ lattice at $U=2$, 6 and 12. At half-filling ($N=9$), $\Omega_{5j}$ is small at weak coupling ($U=2$) and decreases rapidly with increasing $U$. In the strong-coupling limit ($U=12$), the intensity of $\Omega_{5j}$ is nearly quenched, indicating the dominance of localized moments with limited spatial coherence, insufficient to sustain a net altermagnetic response. This reduction is consistent with the suppression of $\langle\Delta_{\mathrm{spin}}\rangle$ observed in Fig. \ref{fig:Fig2}.

In contrast, both one-hole-doped ($N=8$) and one-electron-doped ($N=10$) systems exhibit pronounced and spatially anisotropic patterns characterized by a $d$-wave-like alternating sign structure between neighboring sites. These patterns persist from intermediate ($U=6$) to strong coupling ($U=12$) regimes, indicating that the spatial anisotropy remains robust despite increasing local moment formation. The electron-doped system consistently shows larger $\Omega_{5j}$ amplitudes than the hole-doped system, in agreement with the enhanced $\langle\Delta_{\mathrm{spin}}\rangle$ (Fig. \ref{fig:Fig2}) and reduced local moment $\bar{m}$ (Fig. \ref{fig:Fig3}). Thus, the real-space heatmaps confirm that altermagnetic correlations arise from the combination of geometric frustration and residual charge carrier mobility.

\begin{figure}[h]
\centering
\includegraphics[scale=0.34]{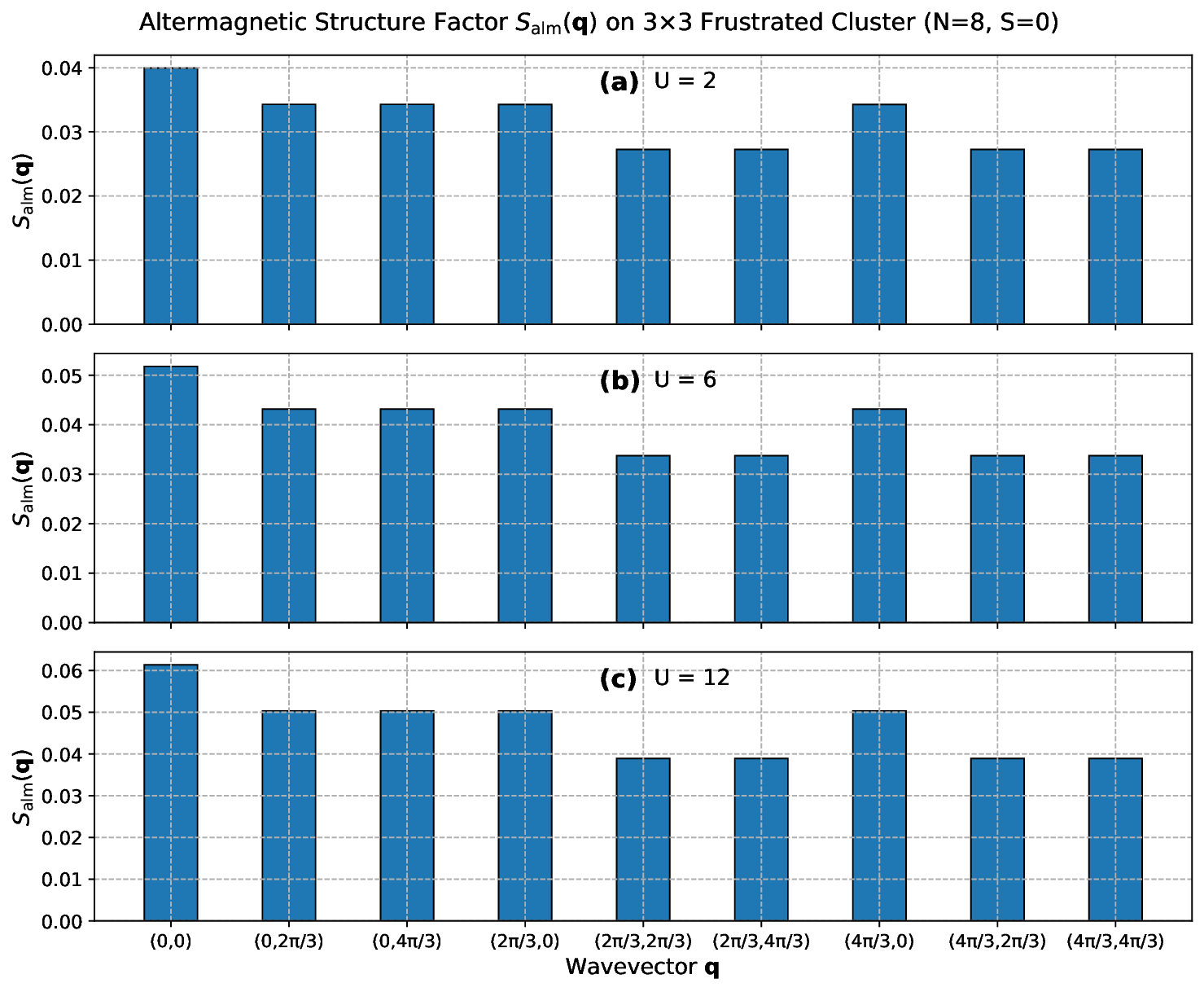}
\caption{Altermagnetic structure factor $S_{\mathrm{alm}}(\mathbf{q})$ for the one-hole-doped ($N=8$) fully frustrated $3\times3$ lattice at $U=2$, 6 and 12.}
\label{fig:Fig5}
\end{figure}

To characterize the momentum-space signature of altermagnetic correlations, we compute the altermagnetic structure factor $S_{\mathrm{alm}}(\mathbf{q})$ [Eq. (\ref{almsf})]. Figure \ref{fig:Fig5} shows $S_{\mathrm{alm}}(\mathbf{q})$ for one-hole-doped ($N=8$, $S=0$) fully frustrated $3\times3$ lattice at $U=2$, 6 and 12. The structure factor remains finite across all allowed momenta and does not exhibit pronounced peaks at any specific ordering wavevector. Instead, its weight is broadly distributed over the discrete momentum points, with relatively larger values near $\mathbf{q}=(0,0)$ and comparable contributions at symmetry-related momenta such as $(2\pi/3,0)$, $(0,2\pi/3)$ and $(2\pi/3,2\pi/3)$. This broad distribution indicates the absence of long-range magnetic order and reflects short-range, anisotropic spin correlations. As $U$ increases, the magnitude of $S_{\mathrm{alm}}(\mathbf{q})$ is enhanced while its momentum dependence remains unchanged. This demonstrates that electronic interactions reinforce the anisotropic correlations without modifying their underlying structure in momentum space, consistent with the monotonic increase of $\langle \Delta_{\mathrm{spin}} \rangle$ (Fig. \ref{fig:Fig2}) and the persistent anisotropic real-space patterns (Fig. \ref{fig:Fig4}).

Qualitatively, the electron-doped ($N=10$) system (as shown in Appendix \ref{almsqv0} in Fig. \ref{fig:Fig18}) exhibits similar momentum distribution with larger magnitude and an overall sign reversal, reflecting particle-hole asymmetry in the correlation structure. In contrast, at half-filling ($N=9$, as shown in Appendix \ref{almsqv0} in Fig. \ref{fig:Fig19}), $S_{\mathrm{alm}}(\mathbf{q})$ is strongly suppressed  with irregular signs. These observations establish that geometric frustration and carrier doping together produce robust momentum-dependent spin correlations, whose strength is further reinforced by increasing $U$.

\begin{figure}[h]
\centering
\includegraphics[scale=0.62]{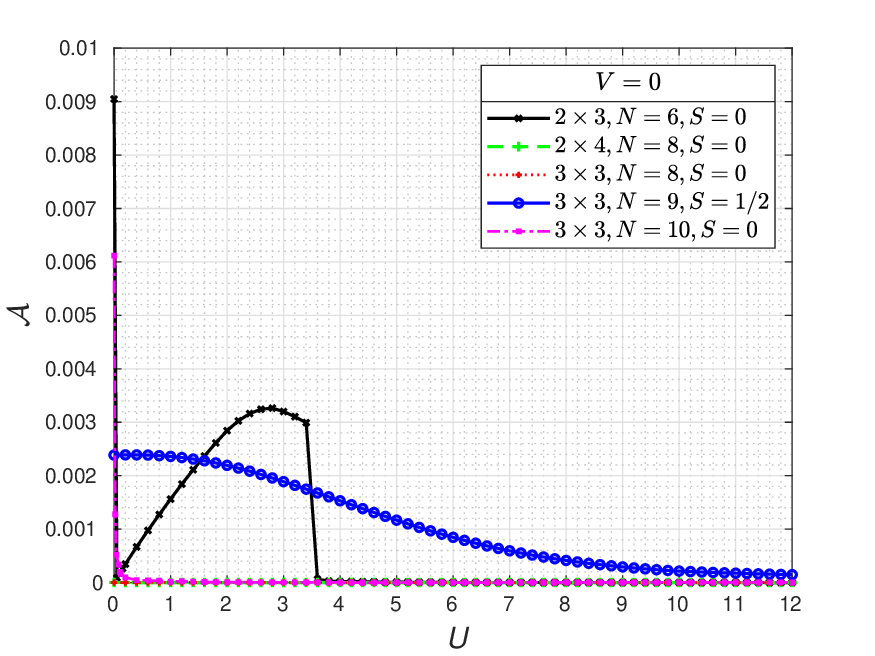}
\caption{Anisotropy parameter $\mathcal{A}$ as a function of $U$ for different lattice geometries and electron filling in the simple Hubbard model ($V=0$). A finite anisotropy appears only in systems with degenerate ground states ($N=9$ on $3\times3$ lattice and $N=6$ on $2\times3$ lattice), while systems with non-degenerate ground states remain isotropic. The suppression of $\mathcal{A}$ in $2\times3$ lattice indicates crossover to a localized regime.}
\label{fig:Fig6}
\end{figure}

To quantify directional anisotropy of the spin correlations, we compute the anisotropy parameter $\mathcal{A}$, which measures the differential spin-response between $x$ and $y$ lattice directions. The results are summarized in Fig. \ref{fig:Fig6}. The trends in $\mathcal{A}$ reveal a direct link between finite anisotropy and ground-state degeneracy. For systems with non-degenerate ground-state --- the doped $3\times3$ lattice ($N=8,10$) and the unfrustrated $2\times4$ lattice ($N=8$) --- $\mathcal{A}$ remains zero across the entire interaction range, indicating that a unique ground state preserves full rotational symmetry between $x$ and $y$ lattice directions. In contrast, a finite anisotropy appears in systems where the ground state is degenerate. For the fully frustrated $3\times3$ lattice at half-filling ($N=9$), which hosts a four-fold degenerate ground state (as shown in Appendix \ref{degenv0} in Table \ref{tab:degen3x3v0}), $\mathcal{A}$ is finite at weak coupling and gradually decreases with increasing $U$. This indicates that geometric frustration enables directionally differentiated spin correlations, but the anisotropy weakens as the system localizes, consistent with the suppression of $\langle \Delta_{\mathrm{spin}} \rangle$ and the rise of $\bar{m}$ (Figs. \ref{fig:Fig2} and \ref{fig:Fig3}). For the partially frustrated $2\times3$ lattice at half-filling ($N=6$), the ground-state is two-fold degenerate for $U\lesssim 3.4$ (as shown in Appendix \ref{degenv0} in Table \ref{tab:degen2x3v0}). In this case, $\mathcal{A}$ initially increases with $U$, peaks near $U\approx3.4$ and then drops to zero at $U_c \approx 3.6$. This sharp collapse coincides with the trends seen in $\langle \Delta_{\mathrm{spin}} \rangle$ (Fig. \ref{fig:Fig2}) and $\bar{m}$ (Fig. \ref{fig:Fig3}), signaling crossover to a localized phase that restores isotropic spin correlations. Together, these results show that on finite clusters, a nonzero anisotropy parameter requires both geometric frustration and ground-state degeneracy: frustration creates the anisotropic correlations, while degeneracy allows them to manifest as a static directional imbalance. 

\begin{figure}[h]
\centering
\includegraphics[scale=0.50]{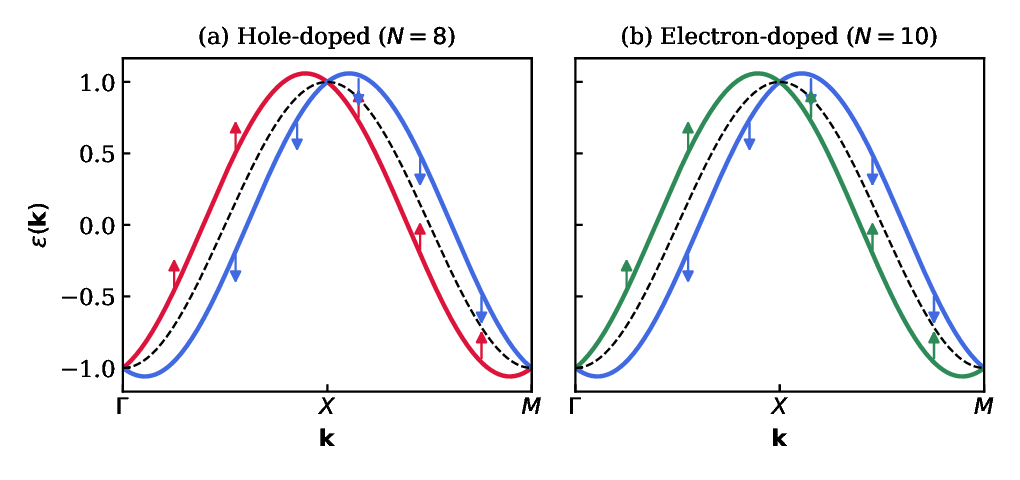}
\caption{Momentum-resolved spin-dependent energy levels $\epsilon(\mathbf{k})$ for the fully frustrated $3\times3$ lattice along the $\Gamma$-$X$-$M$ path for (a) one-hole-doped ($N=8$) and (b) one-electron-doped ($N=10$) systems. Spin degeneracy is lifted in a momentum-dependent manner, with distinct ordering in the two doping cases. Dashed lines denote the non-interacting dispersion.}
\label{fig:Fig7}
\end{figure}

The momentum-resolved spin-dependent energy levels $\epsilon(\mathbf{k})$, shown in Fig. \ref{fig:Fig7}, provide a direct reciprocal-space signature of altermagnetic correlations. In the one-hole-doped system ($N=8$), spin degeneracy is lifted in a momentum-dependent manner: the splitting vanishes at high-symmetry points such as $\Gamma$ and $M$, but becomes finite along intermediate segments of the path, consistent with the short-range anisotropic correlations observed in real space (Fig. \ref{fig:Fig4}) and the broad momentum distribution of $S_{\mathrm{alm}}(\mathbf{q})$ (Fig. \ref{fig:Fig5}). The electron-doped system ($N=10$) shows a similar splitting but with reversed spin ordering, mirroring the sign change in the structure factor (as shown in Appendix \ref{almsqv0} in Fig. \ref{fig:Fig18}) and confirming particle-hole asymmetry. Thus, the interaction-driven anisotropic correlations manifest as a momentum-dependent lifting of spin degeneracy, with the sign structure encoding the asymmetry between electron and hole doping.

Taken together, these results demonstrate that geometric frustration and carrier doping play complementary roles in stabilizing altermagnetic correlations in the simple Hubbard model. While frustration is essential to suppress conventional N\'eel order and enable anisotropic spin textures, the introduction of mobile charge carriers induces robust compensated altermagnetic order that is absent on the unfrustrated lattice. The response is maximized in the intermediate-coupling regime where local moments coexist with sufficient itinerancy, as consistently reflected in real-space patterns, momentum-space structure factor, band splitting and the anisotropy parameter. A pronounced particle-hole asymmetry, with stronger response for electron doping, further underscores the sensitivity of this mechanism to the frustrated band structure. These findings establish a microscopic baseline for interaction-driven altermagnetism before examining the effects of nearest-neighbor interactions in the extended Hubbard model.

\subsection{Extended Hubbard Model}
\label{ehm}
Building on the results of the simple Hubbard model, we now examine the role of nonlocal Coulomb interactions within the extended Hubbard model. In addition to the on-site interaction $U$, the nearest-neighbor (NN) Coulomb repulsion $V$ introduces competing charge fluctuations that modifies the spin correlations, particularly in low-dimensional and geometrically frustrated systems. We consider two representative interaction strengths, $U=4$ and $U=10$, and vary $V$ in the range $0 < V \le U/2$. This regime, motivated by earlier studies \cite{Callaway1990, Hanna2017}, captures the onset of charge-ordering tendencies without dominating over local moment formation.
\subsubsection{U=4 case}
\label{ehmu4}
\begin{figure}[h]
\centering
\includegraphics[scale=0.62]{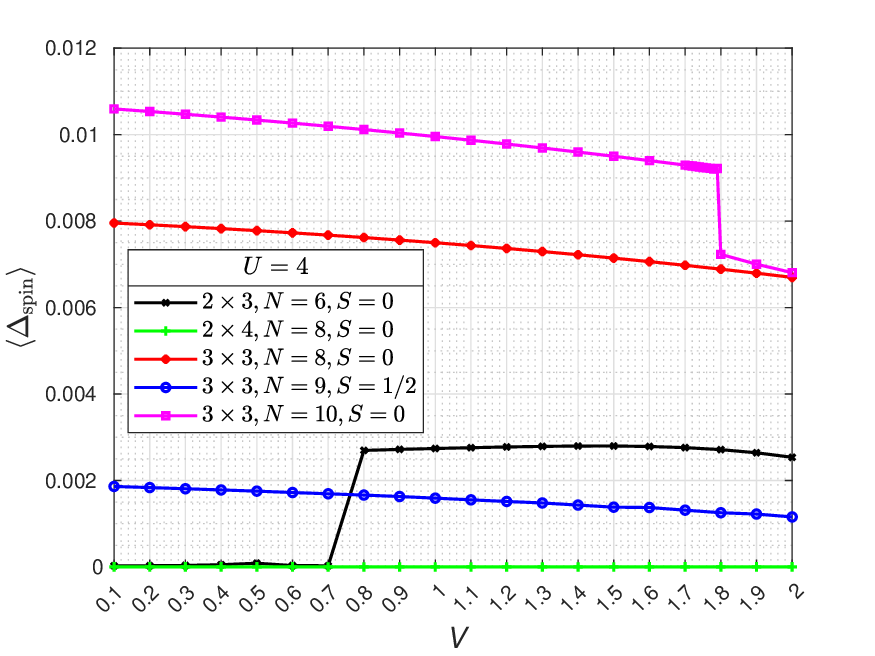}
\caption{Average altermagnetic spin-response $\langle \Delta_{\mathrm{spin}} \rangle$ versus nearest-neighbor Coulomb interaction $V$ at $U=4$.}
\label{fig:Fig8}
\end{figure}

The effect of NN Coulomb interaction $V$ on the average altermagnetic spin-response $\langle \Delta_{\mathrm{spin}} \rangle$ at $U=4$ is examined in Fig. \ref{fig:Fig8}. For the unfrustrated $2\times4$ lattice at half-filling ($N=8$), $\langle \Delta_{\mathrm{spin}} \rangle$ remains zero for all values of $V$, implying that in the absence of geometric frustration, NN interactions are insufficient to drive anisotropic spin correlations. For the fully frustrated $3\times3$ lattice at half-filling ($N=9, S=1/2$), $\langle \Delta_{\mathrm{spin}} \rangle$ decreases gradually with $V$, indicating a continuous suppression of the already weak anisotropic correlations by NN repulsion. In the doped $3\times3$ lattice, $\langle \Delta_{\mathrm{spin}} \rangle$ decreases monotonically with a clear contrast between hole and electron doping: the hole-doped system ($N=8$) shows a weak monotonic decrease, while the electron-doped system ($N=10$) undergoes a gradual reduction followed by a sharp drop near $V \approx 1.8$, signaling a qualitative change in the underlying correlation structure. In the partially frustrated $2\times3$ lattice at half-filling ($N=6$), a qualitatively different trend emerges. Here, $\langle \Delta_{\mathrm{spin}} \rangle$ remains zero for small $V$ and rises abruptly above a threshold $V \approx 0.7$, beyond which it saturates at a finite value. These results reflect that the effect of $V$ is geometry- and doping-dependent: it gradually suppresses the altermagnetic spin-response on the fully frustrated $3\times3$ torus, induces it on the partially frustrated $2\times3$ cylinder, and has no effect on the unfrustrated $2\times4$ cylindrical lattice. 

\begin{figure}[h]
\centering
\includegraphics[scale=0.62]{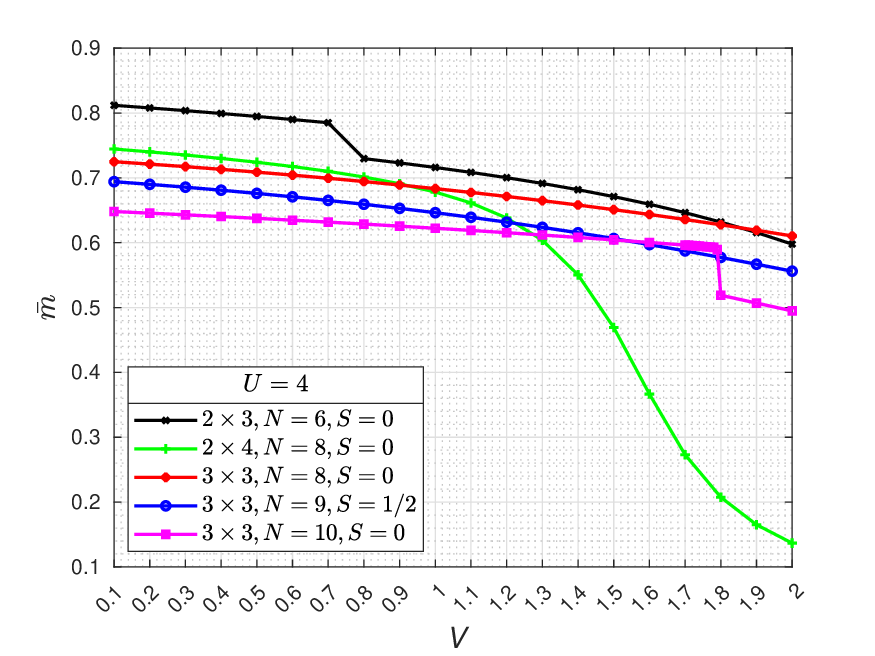}
\caption{Average local moment $\bar{m}$ as a function of nearest-neighbor interaction $V$ at $U=4$. The local moment decreases with increasing $V$ in all cases, with sharp discontinuity near $V\approx 0.7$ in the partially frustrated $2\times3$ lattice and near $V\approx 1.8$ in the electron-doped $3\times3$ lattice.}
\label{fig:Fig9}
\end{figure}

To clarify the influence of NN interaction on electron localization, Fig. \ref{fig:Fig9} shows the average local moment $\bar{m}$ as a function of $V$ at $U=4$. Across all considered geometries, $\bar{m}$ decreases with increasing $V$, indicating that inter-site repulsion competes with on-site repulsion and suppresses local moment formation. For the fully frustrated $3\times3$ lattice at half-filling ($N=9$, $S=1/2$), $\bar{m}$ decreases gradually with $V$, consistent with the simultaneous suppression of $\langle \Delta_{\mathrm{spin}} \rangle$ observed in Fig. \ref{fig:Fig8}, confirming that the weakening of anisotropic correlations accompanies reduced local moment amplitude. In the doped cases, the hole-doped system ($N=8$) shows a moderate reduction, while the electron-doped system ($N=10$) displays a gradual decrease followed by a sharp drop near $V\approx1.8$, coinciding with the discontinuity in $\langle \Delta_{\mathrm{spin}} \rangle$ (Fig. \ref{fig:Fig8}).

The partially frustrated $2\times3$ lattice at half-filling ($N=6$) shows a weak dependence at small $V$ but exhibits a noticeable variation near $V\approx0.7$, coinciding with the onset of a finite altermagnetic response in Fig. \ref{fig:Fig8}. By contrast, the unfrustrated $2\times4$ lattice at half-filling ($N=8$) exhibits a sharp decrease of $\bar{m}$ beyond $V\gtrsim1.2$, while $\langle \Delta_{\mathrm{spin}} \rangle$ remains zero throughout (Fig. \ref{fig:Fig8}), indicating that reduced localization alone is insufficient to induce altermagnetic correlations without geometric frustration.

\begin{figure}[h]
\centering
\includegraphics[scale=0.34]{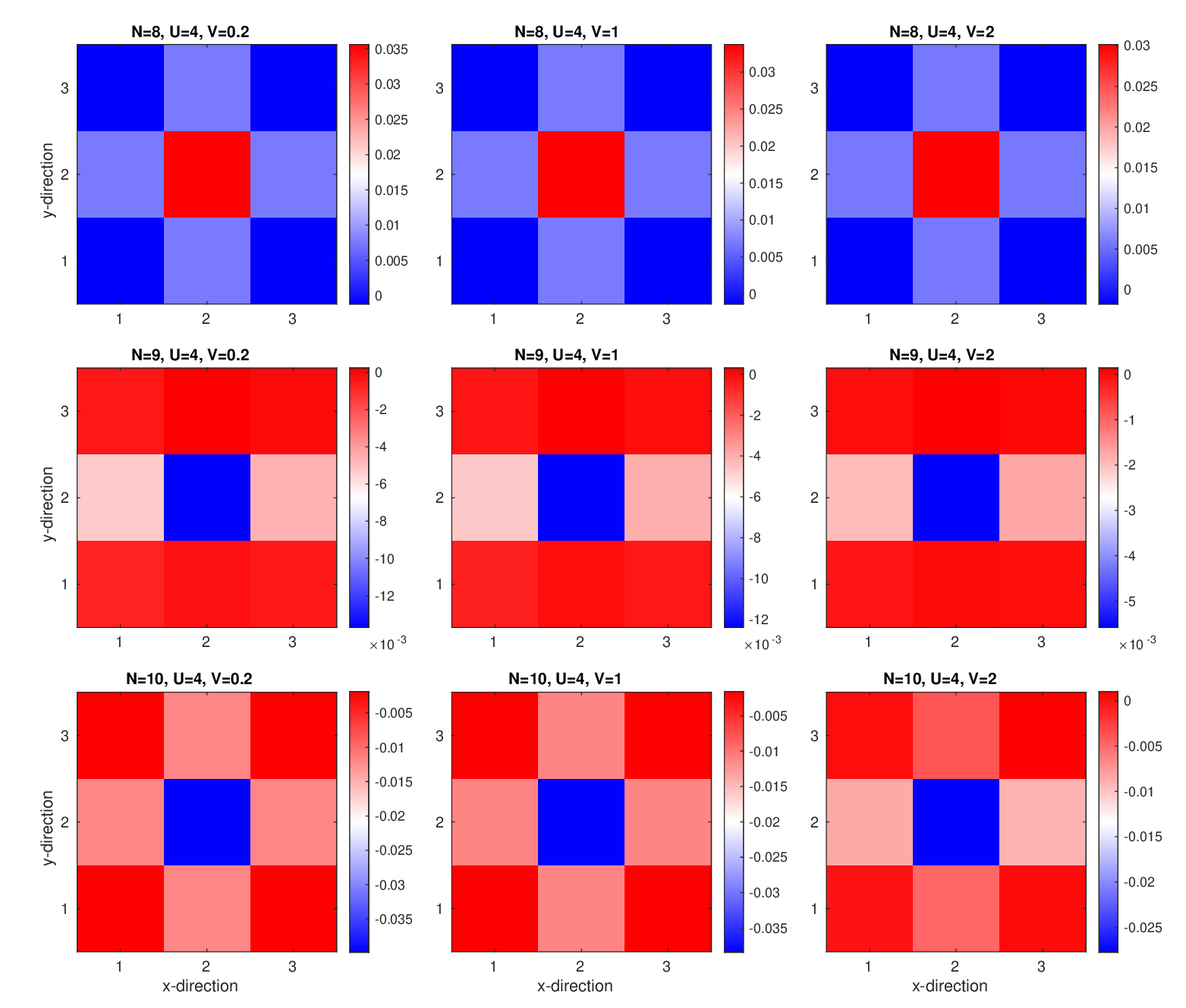}
\caption{Real-space heatmaps of the altermagnetic correlation matrix elements $\Omega_{5j}$ relative to the central site ($i=5$) for the fully frustrated $3\times3$ lattice at $U=4$. Rows correspond to one-hole doping ($N=8$, top), half-filling ($N=9$, middle), and one-electron doping ($N=10$, bottom), while columns show $V=0.2$, 1 and 2.}
\label{fig:Fig10}
\end{figure}

Figure \ref{fig:Fig10} presents real-space heatmaps of the altermagnetic correlation matrix elements $\Omega_{5j}$ for the fully frustrated $3\times3$ lattice at $U=4$ for representative NN interaction strengths $V=0.2$, 1, and 2. For the doped systems ($N=8,10$), the alternating sign pattern characteristic of altermagnetic correlations remains visible across all $V$, indicating that the directional structure is robust against NN repulsion. In the hole-doped system ($N=8$), the magnitude of $\Omega_{5j}$ decreases slightly as $V$ increases from $0.2$ to $2$, consistent with the weak reduction of $\langle\Delta_{\mathrm{spin}}\rangle$ in Fig. \ref{fig:Fig8}. In the electron-doped system ($N=10$), the correlations are strong at small $V$ but become substantially reduced at $V=2$, mirroring the sharp drop in $\langle\Delta_{\mathrm{spin}}\rangle$ (Fig. \ref{fig:Fig8}) and the reduction of $\bar{m}$ (Fig. \ref{fig:Fig9}). A sign reversal between $N=8$ and $N=10$ persists across all $V$, reflecting particle-hole asymmetry.

For the half-filled case ($N=9$), the $\Omega_{5j}$ amplitudes are smaller than in the doped systems and decrease further with increasing $V$, consistent with the suppression of $\langle\Delta_{\mathrm{spin}}\rangle$ and $\bar{m}$. Thus, while spatial anisotropy is preserved in the doped systems, its magnitude declines with increasing $V$, with the strongest suppression occurring in the electron-doped and half-filled cases. These real-space observations confirm that NN interactions weaken altermagnetic correlations without destroying their directional character in the presence of mobile carriers.

\begin{figure}[h]
\centering
\includegraphics[scale=0.34]{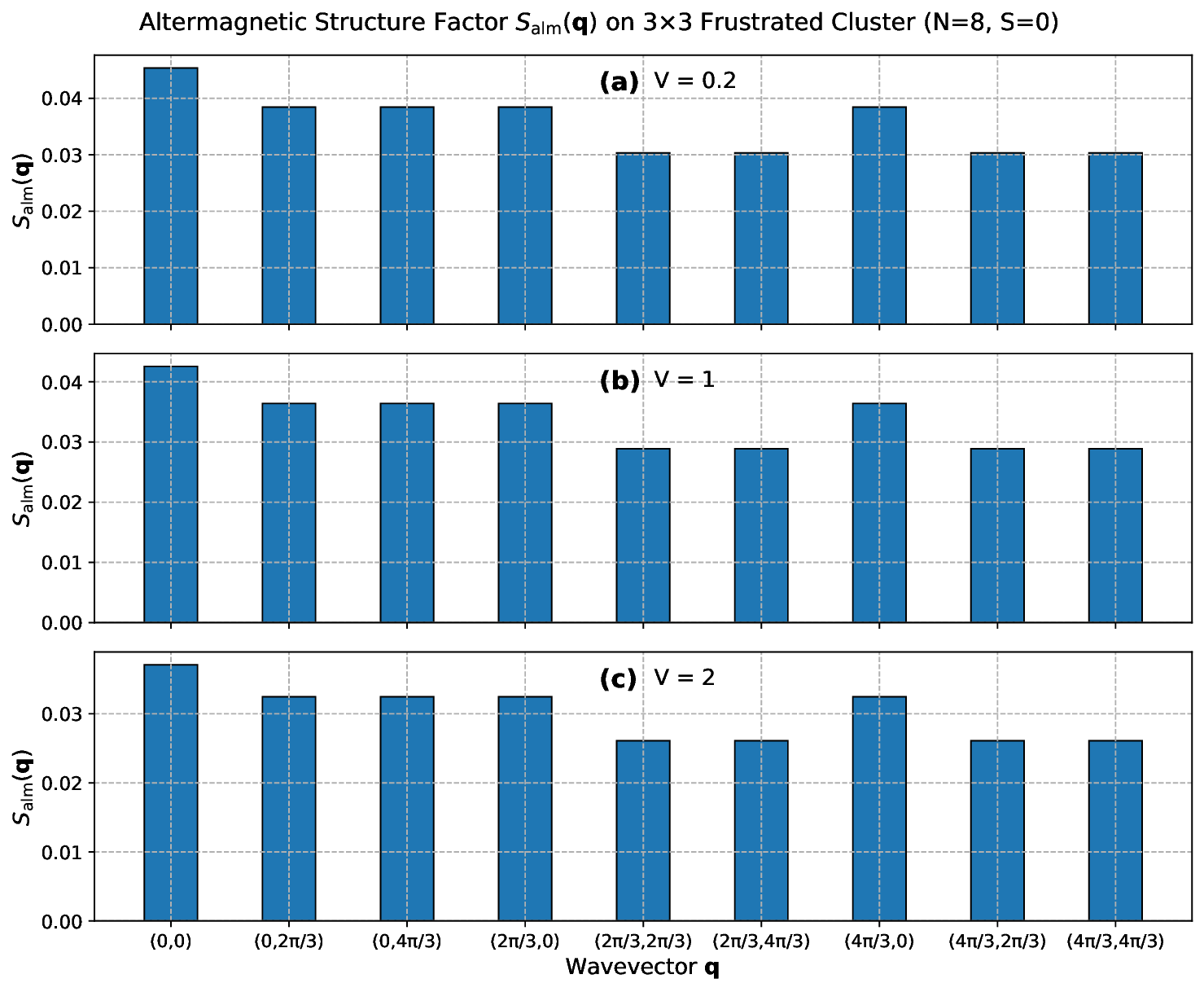}
\caption{Altermagnetic structure factor $S_{\mathrm{alm}}(\mathbf{q})$ for the one-hole-doped ($N=8$) fully frustrated $3\times3$ lattice at $U=4$ and representative nearest-neighbor interactions $V=0.2$, 1 and 2.}
\label{fig:Fig11}
\end{figure}

To examine the momentum-space structure of altermagnetic correlations in presence of NN interaction, Fig. \ref{fig:Fig11} shows the altermagnetic structure factor $S_{\mathrm{alm}}(\mathbf{q})$ for the one-hole-doped ($N=8$) fully frustrated $3\times3$ lattice at $U=4$ and representative values of $V$. The structure factor remains finite across all allowed momenta and exhibits a broad distribution without pronounced peaks, indicating the absence of long-range magnetic order. As $V$ increases, the overall magnitude of $S_{\mathrm{alm}}(\mathbf{q})$ decreases while its momentum distribution remains unchanged. This demonstrates that NN interaction weakens the anisotropic spin correlations without altering their underlying momentum-space structure, consistent with the gradual suppression of $\langle \Delta_{\mathrm{spin}} \rangle$ (Fig. \ref{fig:Fig8}) and the reduced real-space correlation amplitudes observed in Fig. \ref{fig:Fig10}.

The one-electron-doped system ($N=10$, shown in Appendix \ref{almsqu4v} in Fig. \ref{fig:Fig20}) exhibits a similar momentum distribution but with an overall sign reversal and a stronger suppression at large $V$, reflecting persistent particle-hole asymmetry. In contrast, the half-filled system ($N=9$, shown in Appendix \ref{almsqu4v} in Fig. \ref{fig:Fig21}) displays a substantially weak and irregular momentum-space response, indicating the absence of coherent anisotropic spin correlations. These observations confirm that $V$ primarily reduces the amplitude of doping-induced altermagnetic correlations while preserving their momentum-space structure.

\begin{figure}[h]
\centering
\includegraphics[scale=0.62]{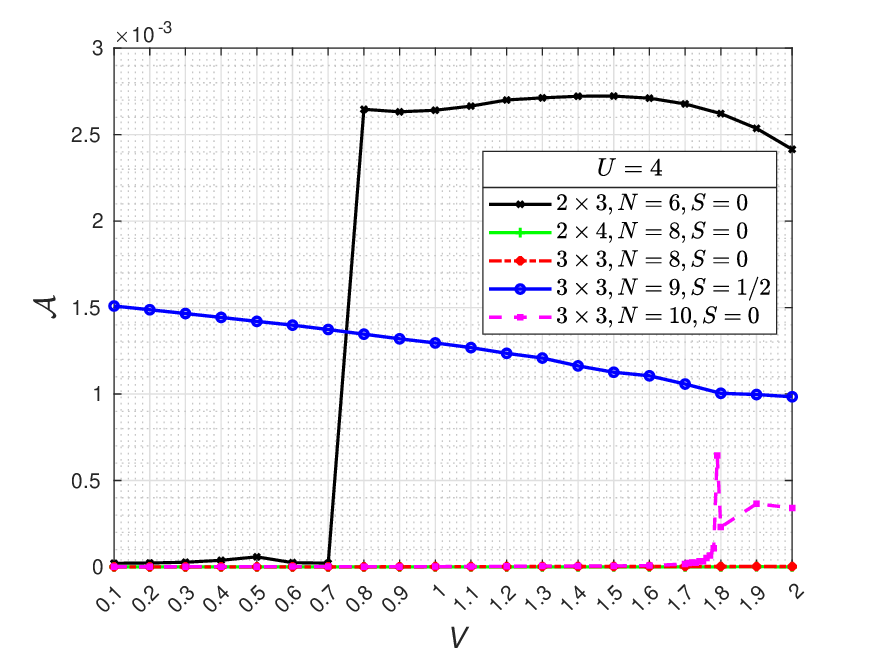}
\caption{Anisotropy parameter $\mathcal{A}$ as a function of nearest-neighbor interaction $V$ at $U=4$ for different lattice geometries and fillings. Finite anisotropy appears only in sectors with degenerate ground states, while non-degenerate sectors remain isotropic.}
\label{fig:Fig12}
\end{figure}

The anisotropy parameter $\mathcal{A}$ as a function of $V$ at $U=4$ is shown in Fig. \ref{fig:Fig12}. The results reveal a close correspondence between directional anisotropy and ground-state degeneracy. For the unfrustrated $2\times4$ lattice at half-filling ($N=8$) and the one-hole-doped $3\times3$ lattice ($N=8$), where the ground-state remains non-degenerate, $\mathcal{A}$ is zero throughout the entire range of $V$. This indicates that a unique ground-state preserves the equivalence between $x$ and $y$ lattice directions, preventing the emergence of a static directional imbalance even when finite altermagnetic correlations are present. In contrast, the fully frustrated $3\times3$ lattice at half-filling ($N=9$), for which the ground-state remains four-fold degenerate, exhibits a finite $\mathcal{A}$ that decreases gradually with increasing $V$, consistent with the suppression of $\langle \Delta_{\mathrm{spin}} \rangle$ and the weakening of anisotropic spin correlations.

A similar interplay between anisotropy and degeneracy is observed in the partially frustrated $2\times3$ lattice ($N=6$) and the one-electron-doped $3\times3$ lattice ($N=10$). For the $2\times3$ lattice, $\mathcal{A}$ remains nearly zero for $V\lesssim0.7$ and increases abruptly once the ground-state becomes degenerate (as shown in Appendix \ref{degenu4v} and Table \ref{tab:degen2x3u4v}), coinciding with the onset of a finite $\langle \Delta_{\mathrm{spin}} \rangle$. In the one-electron-doped $3\times3$ lattice, $\mathcal{A}$ remains negligible up to $V\approx1.7$ but rises sharply near $V\approx1.8$, where the ground-state acquires degeneracy (as shown in Appendix \ref{degenu4v} and Table \ref{tab:degen3x3n10u4v}) and $\langle \Delta_{\mathrm{spin}} \rangle$ simultaneously exhibits a discontinuous change. These results establish that while geometric frustration provides the underlying framework for anisotropic spin correlations, ground-state degeneracy is the essential condition for these correlations to develop into a measurable directional imbalance in finite-size systems.

Collectively, the results at intermediate coupling $U=4$ demonstrate that nearest-neighbor repulsion $V$ modulates altermagnetic correlations in geometry- and doping-dependent manner. On the fully frustrated $3\times3$ lattice, $V$ suppresses the doping-induced altermagnetic spin response, with a sharp instability in the electron-doped system near $V\approx 1.8$. In contrast, moderate $V$ induces finite altermagnetic correlations in the partially frustrated $2\times3$ geometry above a critical value $V\approx 0.7$. The anisotropy parameter $\mathcal{A}$ is finite only in degenerate ground states, highlighting that directional symmetry breaking on finite clusters requires both geometric frustration and ground-state degeneracy. These observations provide essential context for examining the robustness of altermagnetic phases in the strong-coupling regime at $U=10$.

\subsubsection{U=10 case}
\label{ehmu10}
\begin{figure}[h]
\centering
\includegraphics[scale=0.62]{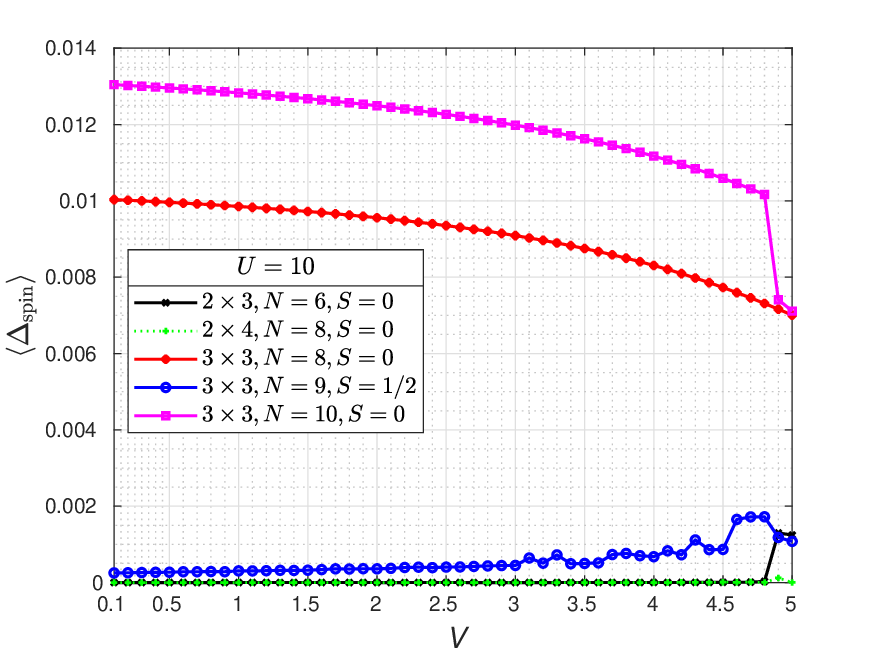}
\caption{Average altermagnetic spin-response $\langle \Delta_{\mathrm{spin}} \rangle$ as a function of nearest-neighbor interaction $V$ at $U=10$ for different lattice geometries and fillings.}
\label{fig:Fig13}
\end{figure}

We now turn to the strong-coupling regime ($U=10$) and examine the effect of NN interaction $V$ on the average altermagnetic spin-response $\langle \Delta_{\mathrm{spin}} \rangle$. Figure \ref{fig:Fig13} shows $\langle \Delta_{\mathrm{spin}} \rangle$ as a function of $V$. For the unfrustrated $2\times4$ lattice at half-filling ($N=8$), $\langle \Delta_{\mathrm{spin}} \rangle$ remains zero for all values of $V$, confirming that geometric frustration remains a prerequisite for altermagnetic correlations even when local moments are fully developed. For the fully frustrated $3\times3$ lattice at half-filling ($N=9$), $\langle \Delta_{\mathrm{spin}} \rangle$ remains small across the entire range of $V$, exhibiting only a weak nonmonotonic variation at intermediate to large $V$. The overall magnitude nevertheless remains low, indicating that anisotropic spin correlations are strongly suppressed at half-filling. Upon doping the fully frustrated $3\times3$ lattice, both the one-hole-doped ($N=8$) and one-electron-doped ($N=10$) systems exhibit finite $\langle\Delta_{\mathrm{spin}}\rangle$ that decreases with increasing $V$, with the electron-doped system undergoing a sharp discontinuity near $V \approx 4.9$, signaling a qualitative reorganization of the underlying correlation structure. The partially frustrated $2\times3$ lattice ($N=6$) displays a qualitatively distinct behavior: $\langle \Delta_{\mathrm{spin}} \rangle$ remains zero for $V\lesssim4.9$ and rises sharply beyond this value, indicating that anisotropic spin correlations in the partially frustrated geometry emerge at sufficiently large NN interaction in the strong-coupling regime. These trends underscore the persistent interplay between geometric frustration, carrier doping and nonlocal interactions in determining the stability of altermagnetic correlations.

\begin{figure}[h]
\centering
\includegraphics[scale=0.62]{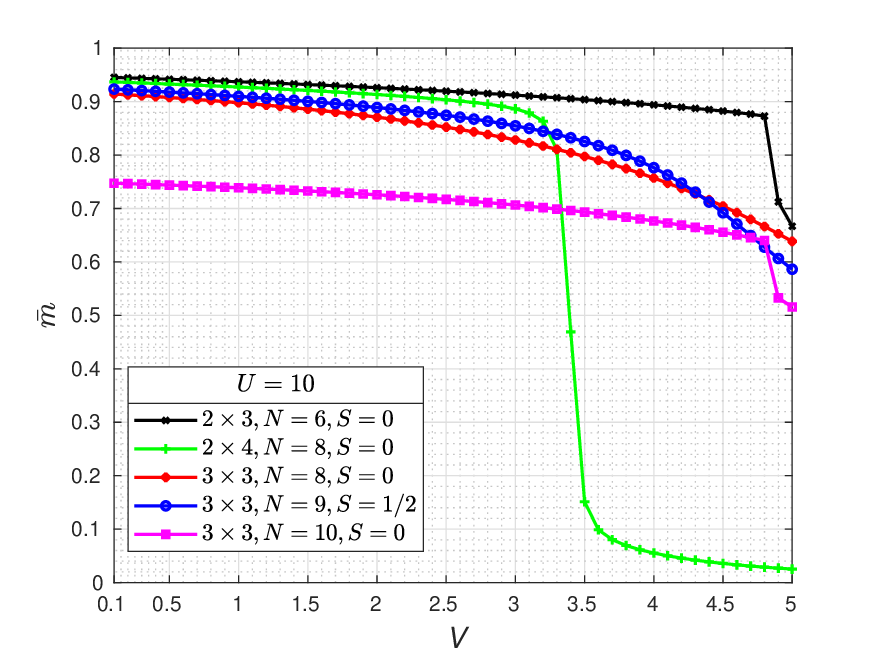}
\caption{Average local moment $\bar{m}$ as a function of nearest-neighbor interaction $V$ at $U=10$. $\bar{m}$ decreases with $V$ in all cases; sharp reductions occur in the one-electron-doped $3\times3$ lattice ($N=10$) and in the partially frustrated $2\times3$ lattice ($N=6$) near $V\approx4.9$.}
\label{fig:Fig14}
\end{figure}

To clarify how NN interactions modify the tendency for electron localization at strong coupling, Fig. \ref{fig:Fig14} shows $\bar{m}$ as a function of $V$ at $U=10$. For the fully frustrated $3\times3$ lattice at half-filling ($N=9$), $\bar{m}$ decreases gradually with $V$, consistent with the small and weakly varying $\langle\Delta_{\mathrm{spin}}\rangle$ observed in Fig. \ref{fig:Fig13}. In the doped $3\times3$ lattice, the one-hole-doped system ($N=8$) exhibits a smooth monotonic reduction of $\bar{m}$, while the one-electron-doped system ($N=10$) shows a pronounced discontinuity near $V\approx4.9$, where $\bar{m}$ decreases sharply, coinciding with the sharp drop in $\langle \Delta_{\mathrm{spin}} \rangle$ (Fig. \ref{fig:Fig13}) and signaling a reorganization of the underlying correlation structure. In the partially frustrated $2\times3$ lattice ($N=6$), a reduction in $\bar{m}$ occurs at large $V$, aligning with the emergence of finite $\langle\Delta_{\mathrm{spin}}\rangle$. In contrast, the unfrustrated $2\times4$ lattice at half-filling ($N=8$) undergoes a strong suppression of $\bar{m}$ beyond $V\approx3.3$, while $\langle \Delta_{\mathrm{spin}} \rangle$ remains zero throughout. These results establish that while NN interactions modify the balance between itinerancy and localization, geometric frustration remains the key ingredient for stabilizing anisotropic spin correlations.

\begin{figure}[h]
\centering
\includegraphics[scale=0.34]{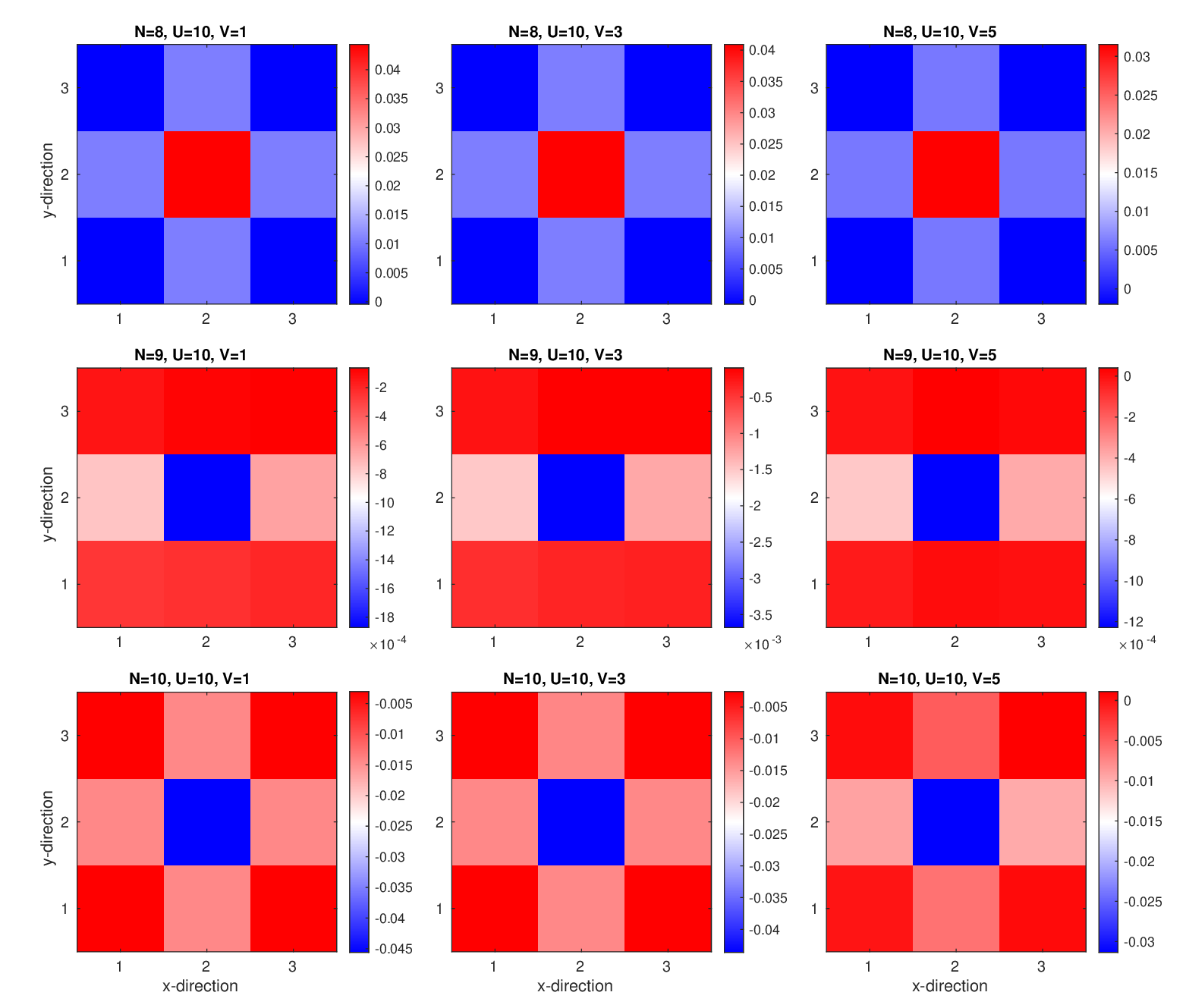}
\caption{Real-space heatmaps of the altermagnetic correlation matrix elements $\Omega_{5j}$ relative to the central site ($i=5$) for the fully frustrated $3\times3$ lattice at $U=10$. Rows correspond to one-hole doping ($N=8$, top), half-filling ($N=9$, middle), and one-electron doping ($N=10$, bottom), while columns show $V=1$, 3 and 5.}
\label{fig:Fig15}
\end{figure}

Figure \ref{fig:Fig15} presents real-space heatmaps of the altermagnetic correlation matrix elements $\Omega_{5j}$ for representative NN interaction strengths $V=1$, 3 and 5 at $U=10$ for the fully frustrated $3\times3$ lattice. In the doped systems ($N=8,10$), the spatial distribution of $\Omega_{5j}$ remains anisotropic across all values of $V$, while preserving the relative sign structure with increasing $V$, indicating that the directional character of the correlations is robust against NN repulsion. For the half-filled system ($N=9$), the spatial pattern remains present but weak, consistent with the suppressed behavior of $\langle\Delta_{\mathrm{spin}}\rangle$ (Fig. \ref{fig:Fig13}) and $\bar{m}$ (Fig. \ref{fig:Fig14}).

\begin{figure}[h]
\centering
\includegraphics[scale=0.34]{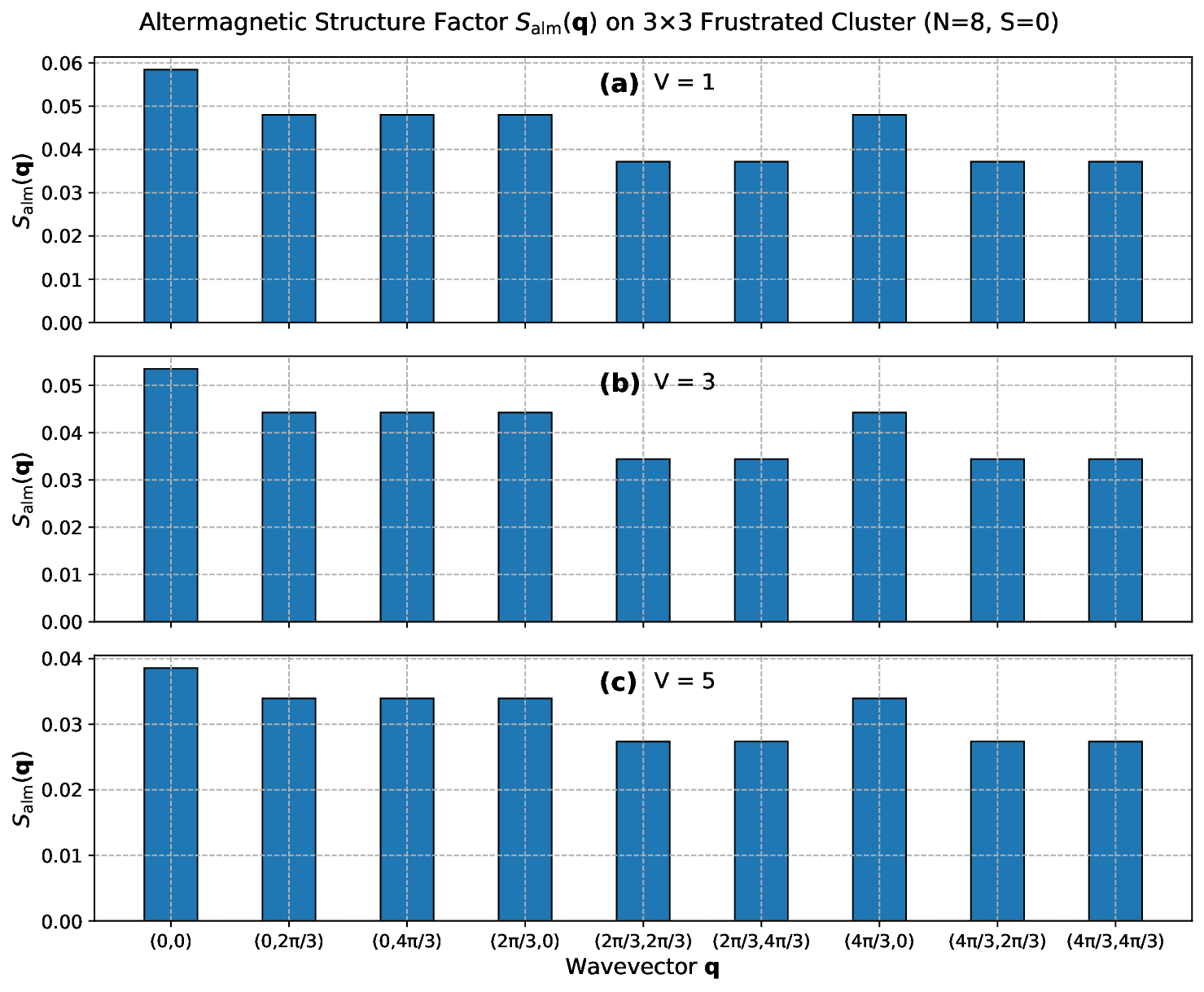}
\caption{Altermagnetic structure factor $S_{\mathrm{alm}}(\mathbf{q})$ for the one-hole-doped ($N=8$) fully frustrated $3\times3$ lattice at $U=10$ and representative NN interactions $V=1$, 3 and 5.}
\label{fig:Fig16}
\end{figure}

To characterize the momentum-space evolution of altermagnetic correlations at strong coupling, we compute the altermagnetic structure factor $S_{\mathrm{alm}}(\mathbf{q})$ for $V=1$, 3 and 5 at $U=10$. Figure \ref{fig:Fig16} shows $S_{\mathrm{alm}}(\mathbf{q})$ for the one-hole-doped ($N=8$) fully frustrated $3\times3$ lattice. The structure factor remains finite across all discrete wavevectors, exhibiting a broad distribution without dominant ordering peaks. As $V$ increases, the magnitude decreases uniformly while the momentum dependence remains unchanged. This indicates that NN interaction reduces the overall amplitude of the altermagnetic correlations without altering their underlying momentum-space structure, consistent with the corresponding reduction in $\langle \Delta_{\mathrm{spin}} \rangle$ (Fig. \ref{fig:Fig13}) and the weakening of real-space correlations (Fig. \ref{fig:Fig15}). Qualitatively, the one-electron-doped system ($N=10$, shown in Appendix \ref{almsqu10v} in Fig. \ref{fig:Fig22}) exhibits a similar broad distribution but with larger magnitude and an overall sign reversal, while at half-filling ($N=9$, shown in Appendix \ref{almsqu10v} in Fig. \ref{fig:Fig23}), $S_{\mathrm{alm}}(\mathbf{q})$ remains strongly suppressed with irregular sign variations.

\begin{figure}[h]
\centering
\includegraphics[scale=0.62]{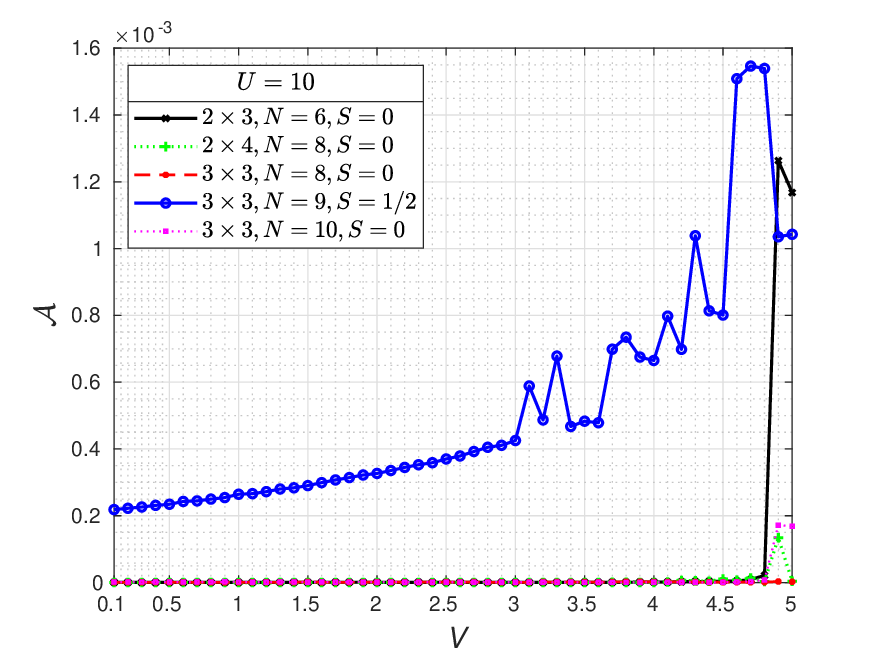}
\caption{Anisotropy parameter $\mathcal{A}$ as a function of nearest-neighbor interaction $V$ at $U=10$. Finite $\mathcal{A}$ appears in systems with degenerate ground states. A sharp increase occurs in the partially frustrated $2\times3$ lattice near $V\approx 4.9$ (onset of two-fold degeneracy) and in the one-electron-doped ($N=10$) fully frustrated $3\times3$ lattice near $V\approx 4.9$ (onset of four-fold degeneracy).}
\label{fig:Fig17}
\end{figure}

Figure \ref{fig:Fig17} shows the anisotropy parameter $\mathcal{A}$ as a function of $V$ at $U=10$. The trends in $\mathcal{A}$ exhibit a direct correspondence with ground-state degeneracy across all lattice geometries. In systems with non-degenerate ground states, including the one-hole-doped $3\times3$ lattice ($N=8$) and the unfrustrated $2\times4$ lattice ($N=8$), $\mathcal{A}$ remains zero throughout the entire range of $V$, indicating that rotational symmetry between $x$ and $y$ lattice directions remains preserved despite finite altermagnetic correlations in the doped frustrated system. By contrast, finite anisotropy emerges only in systems where the ground state becomes degenerate. For the half-filled fully frustrated $3\times3$ lattice ($N=9$), which remains four-fold degenerate, $\mathcal{A}$ remains finite across all $V$. In the partially frustrated $2\times3$ lattice ($N=6$), $\mathcal{A}$ increases sharply near $V\approx4.9$ once the ground state becomes two-fold degenerate (as shown in Appendix \ref{degenu10v} and in Table \ref{tab:degen2x3u10v}). Similarly, in the one-electron-doped $3\times3$ lattice ($N=10$), $\mathcal{A}$ rises steeply near $V\approx4.9$ as the ground-state becomes nearly four-fold degenerate (as shown in Appendix \ref{degenu10v} and in Table \ref{tab:degen3x3n10u10v}), coinciding with the discontinuous changes observed in $\langle\Delta_{\mathrm{spin}}\rangle$. Thus, in finite-size clusters, geometric frustration provides the necessary anisotropic spin correlations, but ground-state degeneracy is the essential condition for these correlations to manifest as a measurable directional imbalance.

In conclusion, the strong-coupling regime at $U=10$ demonstrates that altermagnetism remains robust on geometrically frustrated lattices even when local moments are fully developed. Nearest-neighbor repulsion $V$ suppresses the magnitude of doping-induced correlations while preserving their spatial and momentum-space structure. In specific geometries, however, sufficiently large $V$ can induce the altermagnetic correlations by stabilizing degenerate ground states. These findings highlight that the altermagnetism in the extended Hubbard model is a highly tunable phase, where the interplay of geometric frustration and local and nonlocal Coulomb interactions allows for the precise modulation of directional spin order.

\section{Summary and Conclusion}
\label{summary}
In this work, we have carried out a comprehensive many-body investigation into the emergence of correlation-driven altermagnetism in simple and extended Hubbard model. Using exact diagonalization on finite-size lattice-clusters with varying degrees of geometric frustration, we identify the microscopic conditions necessary for stabilizing compensated anisotropic spin correlations. Motivated by recent advances in altermagnetism \cite{smejkalprx2022, Jungwirth2025, Hussain2025}, symmetry classifications \cite{smejkal2022, Zhou2025, Sike2024} and microscopic modeling \cite{Paul2024, Das2024, Hana2025, Pedro2025, Peng2025}, we addressed a fundamental open question: whether purely electronic correlations on symmetric, geometrically frustrated lattices are sufficient to induce robust altermagnetic correlations in the absence of single-particle anisotropy or spin-orbit coupling. Our results demonstrate that purely electronic correlations on symmetric frustrated lattices are sufficient to induce robust altermagnetic correlations. This mechanism results independently of engineered hopping anisotropy \cite{Mohsen2025, Mohsen2026, Nitin2025} or spin-orbit coupling \cite{Das2024, Ali2025, Ming2025, Costa2025}, suggesting that the confluence of many-body effects and lattice topology provides a minimal framework for breaking directional symmetry in strongly correlated systems.

Investigation of the simple Hubbard model ($V=0$) reveals that altermagnetic correlations are governed by the interplay of geometric frustration, carrier doping and interaction strength. In the fully frustrated $3\times3$ lattice, geometric frustration suppresses altermagnetic correlations at half-filling despite well-developed local moments. However, single-carrier doping (hole or electron) induces a robust enhancement of $\langle\Delta_{\mathrm{spin}}\rangle$, accompanied by anisotropic real-space patterns, broad momentum-space structure factor $S_{\mathrm{alm}}(\mathbf{q})$ and momentum-dependent spin splitting. In the partially frustrated $2\times3$ lattice, $\langle\Delta_{\mathrm{spin}}\rangle$ exhibits a non-monotonic interaction dependence where it collapses near $U_c\approx3.6$, signaling a crossover from an itinerant altermagnetic regime to an isotropic localized phase. This response is absent on the unfrustrated $2\times4$ lattice, establishing geometric frustration as a prerequisite. A defining feature of our findings is the strict correlation between the anisotropy parameter $\mathcal{A}$ and ground-state degeneracy. Across all cases, $\mathcal{A}$ becomes finite only in degenerate ground-state manifolds, demonstrating that while geometric frustration enables anisotropic spin correlations, ground-state degeneracy is required for them to manifest as a static directional imbalance on finite clusters.

The inclusion of nearest-neighbor interaction $V$ in the extended Hubbard model introduces an additional tuning parameter for altermagnetic stability. Across both intermediate ($U=4$) and strong ($U=10$) coupling regimes, increasing $V$ suppresses $\langle\Delta_{\mathrm{spin}}\rangle$ in the fully frustrated $3\times3$ lattice by competing with on-site repulsion, while stabilizing a finite response in the partially frustrated $2\times3$ lattice through the emergence of ground-state degeneracy beyond a critical interaction strength. Real-space and momentum-space analyses further show that although $V$ reduces the overall correlation amplitude, the underlying $d$-wave-like anisotropic structure remains robust. The sharp discontinuities observed at large $V$, particularly in the electron-doped ($N=10$) $3\times3$ lattice, signal a reorganization of the underlying correlation structure driven by nonlocal Coulomb repulsion.

In conclusion, this study establishes a purely correlation-driven route to altermagnetism on symmetric geometrically frustrated lattices. Geometric frustration enables anisotropic spin textures, carrier doping stabilizes them through residual itinerancy and nearest-neighbor interactions provide a tunable parameter that can either suppress or induce the altermagnetic correlation depending on lattice geometry and ground-state degeneracy. These findings broaden the microscopic understanding of altermagnetism and identify strongly correlated frustrated systems as promising platforms for realizing compensated anisotropic magnetism.
\section*{Funding}

The authors declare that no funds, grants, or other financial support were received during the preparation of this manuscript.

\section*{Appendix}
\appendix
\section{Altermagnetic structure factor for one-electron-doped ($N=10$) and half-filled ($N=9$) systems of fully frustrated $3\times 3$ square lattice}
\label{almsq}
\subsection{Simple Hubbard Model: $V=0$}
\label{almsqv0}
Here we present the altermagnetic structure factor $S_{\mathrm{alm}}(\mathbf{q})$ for one-electron-doped ($N=10$) and half-filled ($N=9$) cases of fully frustrated $3\times 3$ square lattice. These results complement the discussion in Sec. \ref{shm} by illustrating the particle-hole asymmetry and the suppression of momentum-space anisotropy at half-filling.
\begin{figure}[h]
\centering
\includegraphics[scale=0.35]{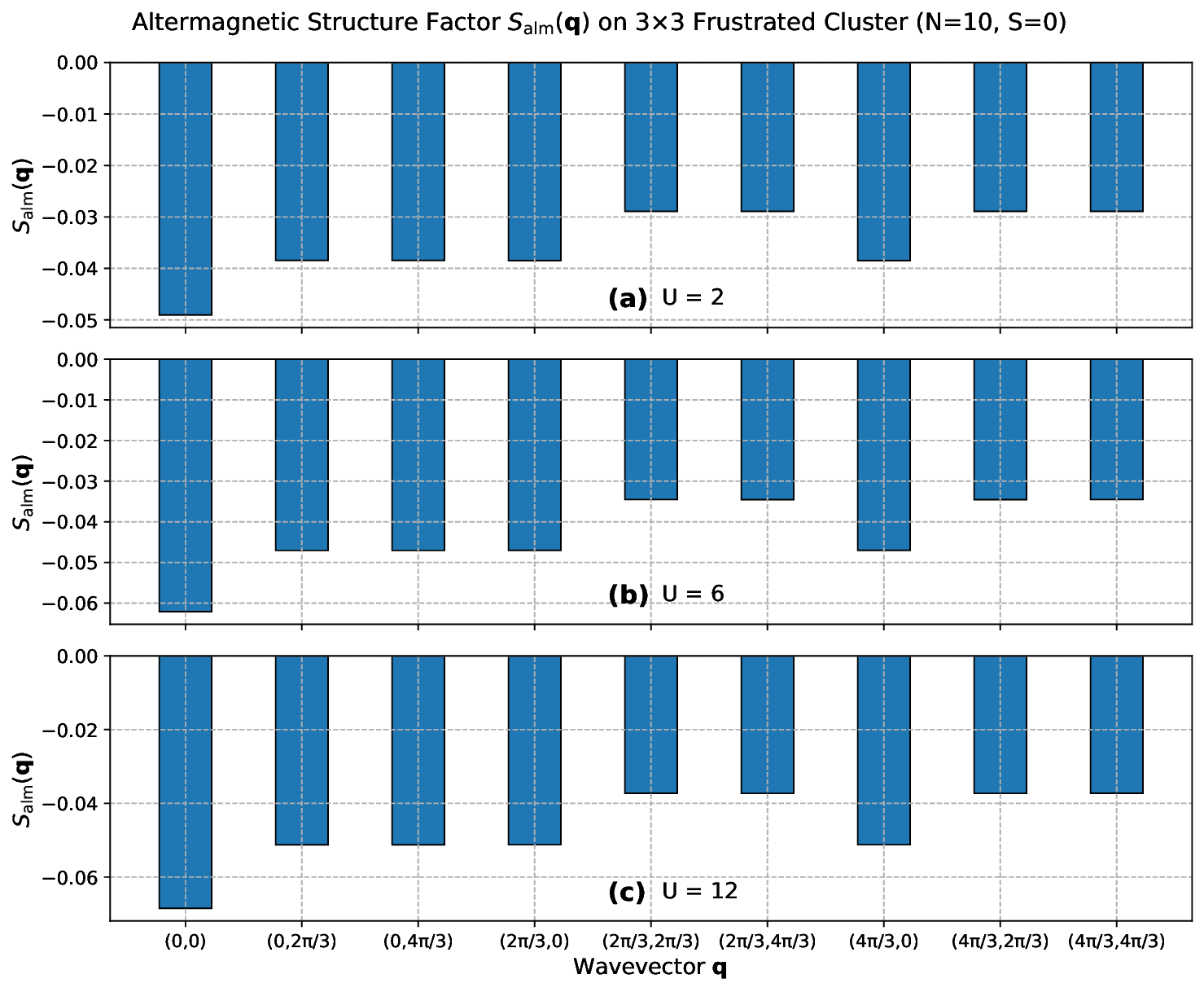}
\caption{Altermagnetic structure factor $S_{\mathrm{alm}}(\mathbf{q})$ for one-electron-doped ($N=10$, $S=0$) case of fully frustrated $3\times3$ square lattice at $U=2$, 6 and 12. The momentum dependence resembles that of one-hole-doped case (Fig. \ref{fig:Fig5}) but with reversed sign and higher magnitude. Increasing $U$ raises the magnitude while preserving the broad distribution without sharp peaks.}
\label{fig:Fig18}
\end{figure}

\begin{figure}[h]
\centering
\includegraphics[scale=0.35]{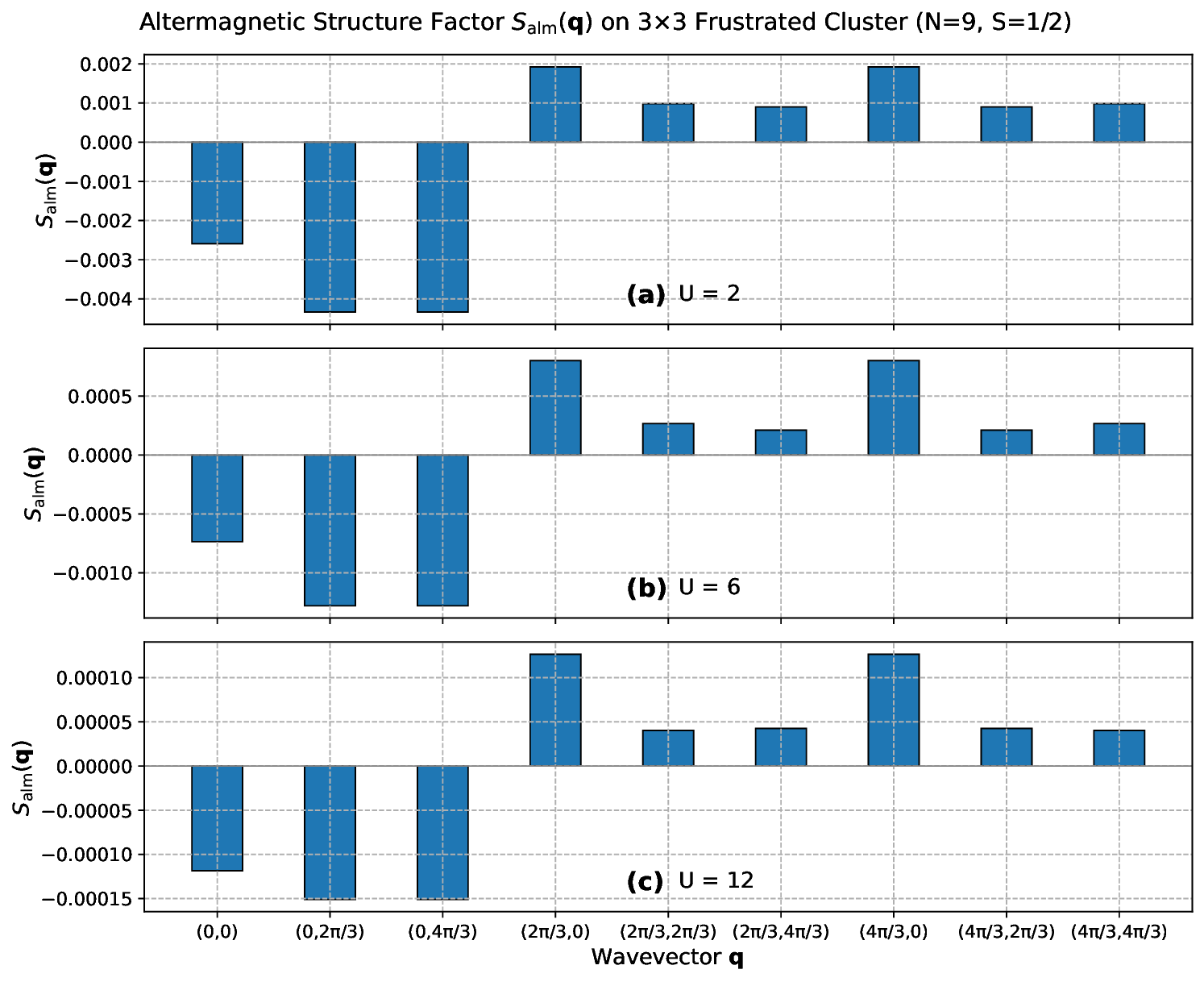}
\caption{Altermagnetic structure factor $S_{\mathrm{alm}}(\mathbf{q})$ for half-filled ($N=9$, $S=1/2$) case of fully frustrated $3\times3$ square lattice at $U=2$, 6 and 12. The structure factor is strongly suppressed across all momenta and exhibits irregular sign variations, indicating the absence of a coherent anisotropic spin structure.}
\label{fig:Fig19}
\end{figure}

\subsection{Extended Hubbard Model: $U=4$}
\label{almsqu4v}
Here we present the altermagnetic structure factor $S_{\mathrm{alm}}(\mathbf{q})$ for the one-electron-doped ($N=10$) and half-filled ($N=9$) systems of the fully frustrated $3\times3$ lattice at $U=4$ and representative NN interactions $V=0.2$, 1 and 2. These results complement the discussion in Sec. \ref{ehmu4} by illustrating the persistence of particle-hole asymmetry and the suppression of momentum-space anisotropy at half-filling in the presence of NN interaction.
\begin{figure}[h]
\centering
\includegraphics[scale=0.35]{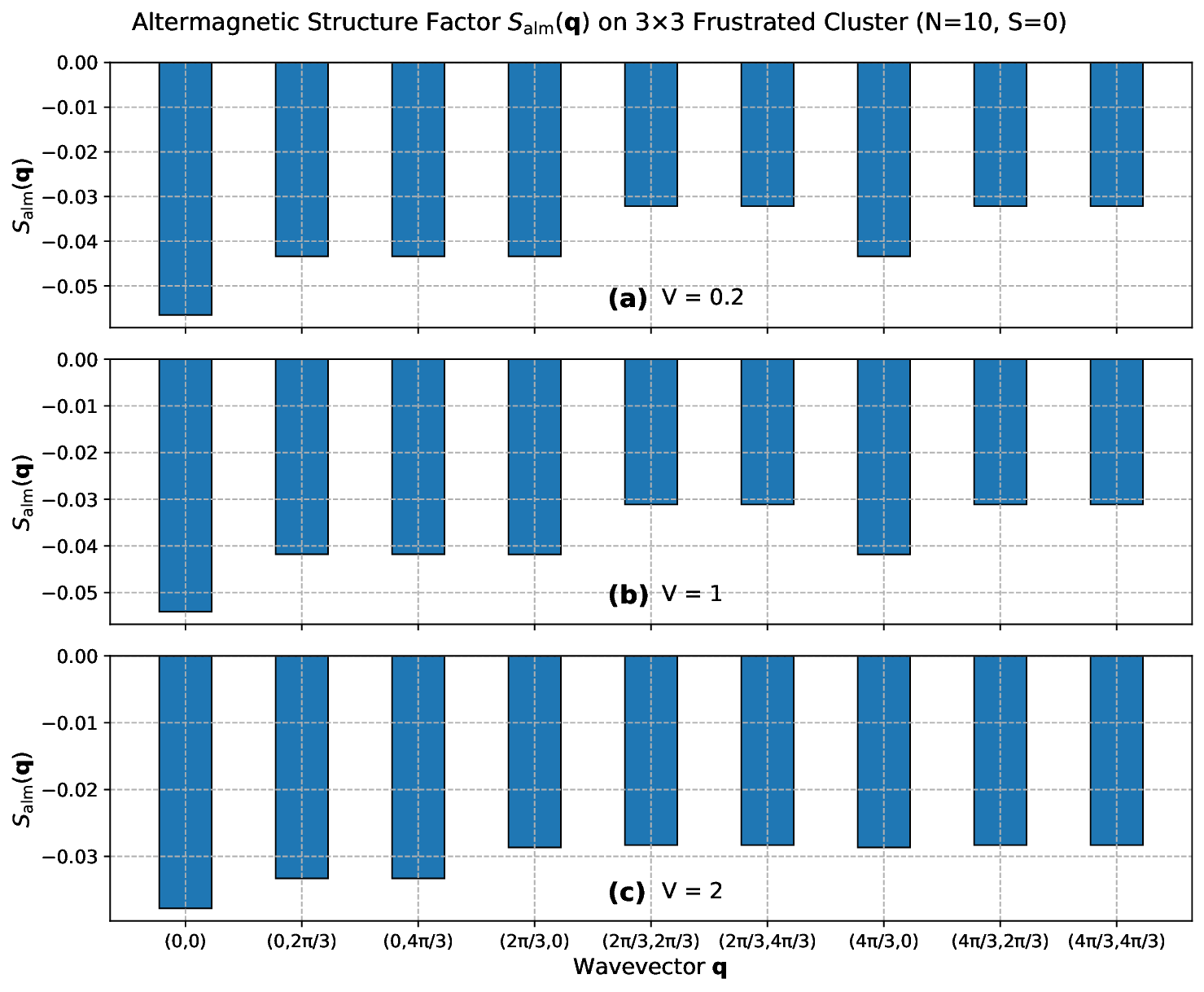}
\caption{Altermagnetic structure factor $S_{\mathrm{alm}}(\mathbf{q})$ for the one-electron-doped ($N=10$) fully frustrated $3\times3$ lattice at $U=4$ and $V=0.2$, 1 and 2. The momentum distribution resembles the one-hole-doped case (Fig. \ref{fig:Fig11}) but with reversed sign and stronger suppression at large $V$.}
\label{fig:Fig20}
\end{figure}

\begin{figure}[h]
\centering
\includegraphics[scale=0.35]{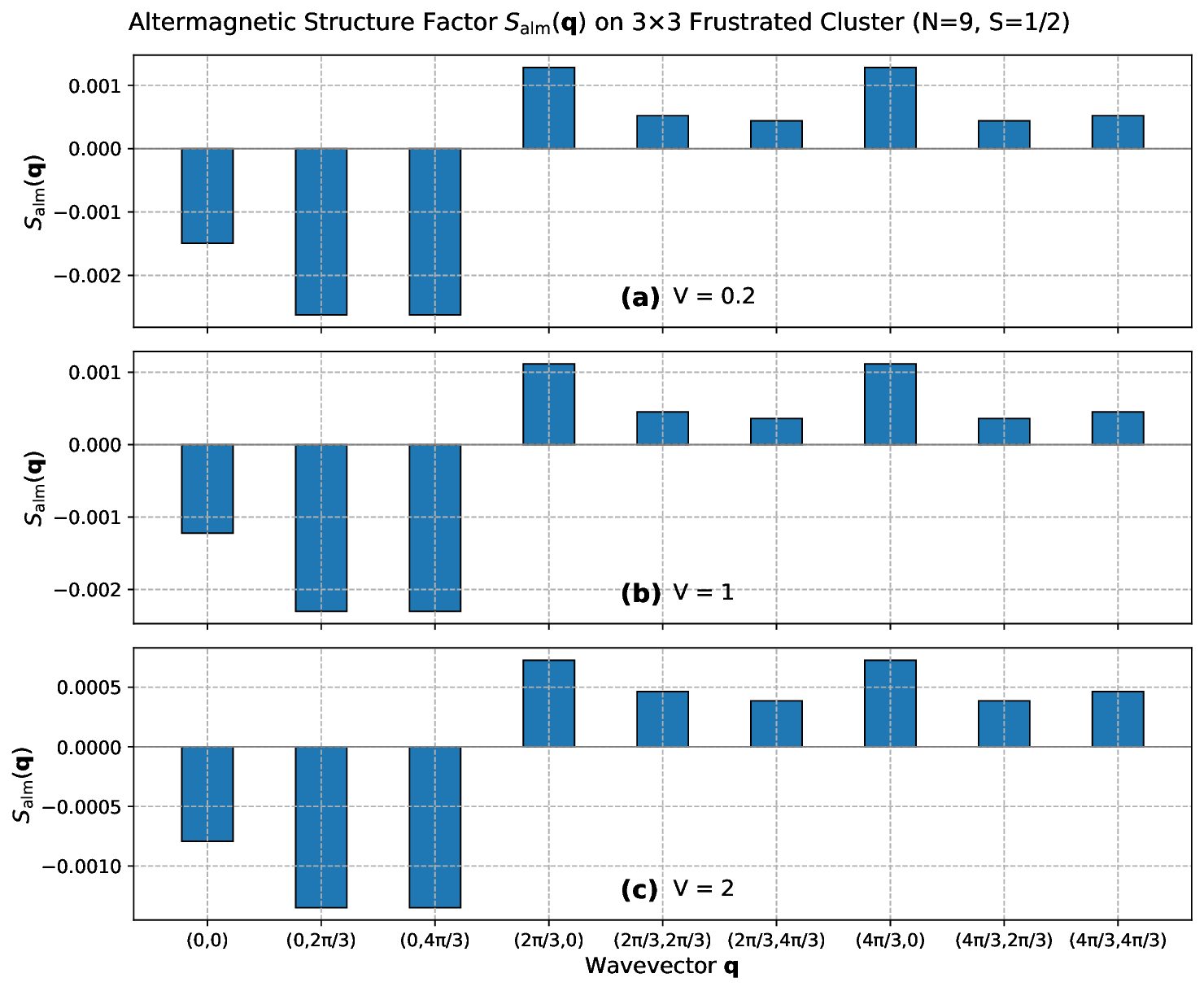}
\caption{Altermagnetic structure factor $S_{\mathrm{alm}}(\mathbf{q})$ for the half-filled ($N=9$, $S=1/2$) fully frustrated $3\times3$ lattice at $U=4$ and $V=0.2$, 1 and 2. The structure factor remains weak across all momenta and exhibits irregular sign variations, indicating the absence of coherent momentum-space anisotropy.}
\label{fig:Fig21}
\end{figure}

\subsection{Extended Hubbard Model: $U=10$}
\label{almsqu10v}

Here we present the altermagnetic structure factor $S_{\mathrm{alm}}(\mathbf{q})$ for the one-electron-doped ($N=10$) and half-filled ($N=9$) fully frustrated $3\times3$ lattice at $U=10$ for representative NN interactions $V=1$, 3 and 5. These results complement the discussion in Sec. \ref{ehmu10} by illustrating the persistence of particle-hole asymmetry and the suppression of momentum-space anisotropy at half-filling in the presence of NN interaction.

\begin{figure}[h]
\centering
\includegraphics[scale=0.35]{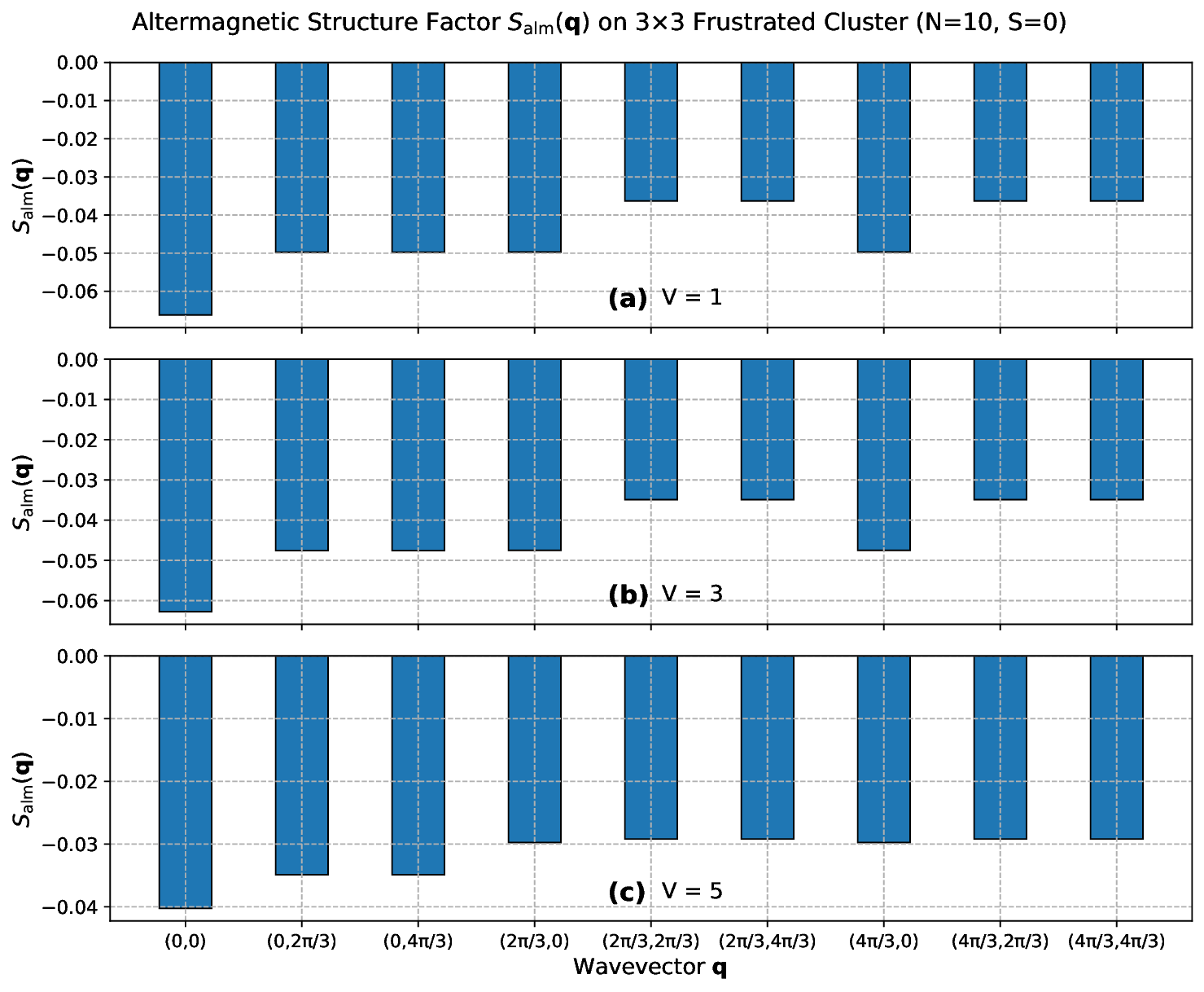}
\caption{Altermagnetic structure factor $S_{\mathrm{alm}}(\mathbf{q})$ for the one-electron-doped ($N=10$) fully frustrated $3\times3$ lattice at $U=10$ and $V=1$, 3 and 5. The momentum distribution resembles the one-hole-doped case shown in Fig. \ref{fig:Fig16}, but with reversed sign and stronger suppression at large $V$.}
\label{fig:Fig22}
\end{figure}

\begin{figure}[h]
\centering
\includegraphics[scale=0.35]{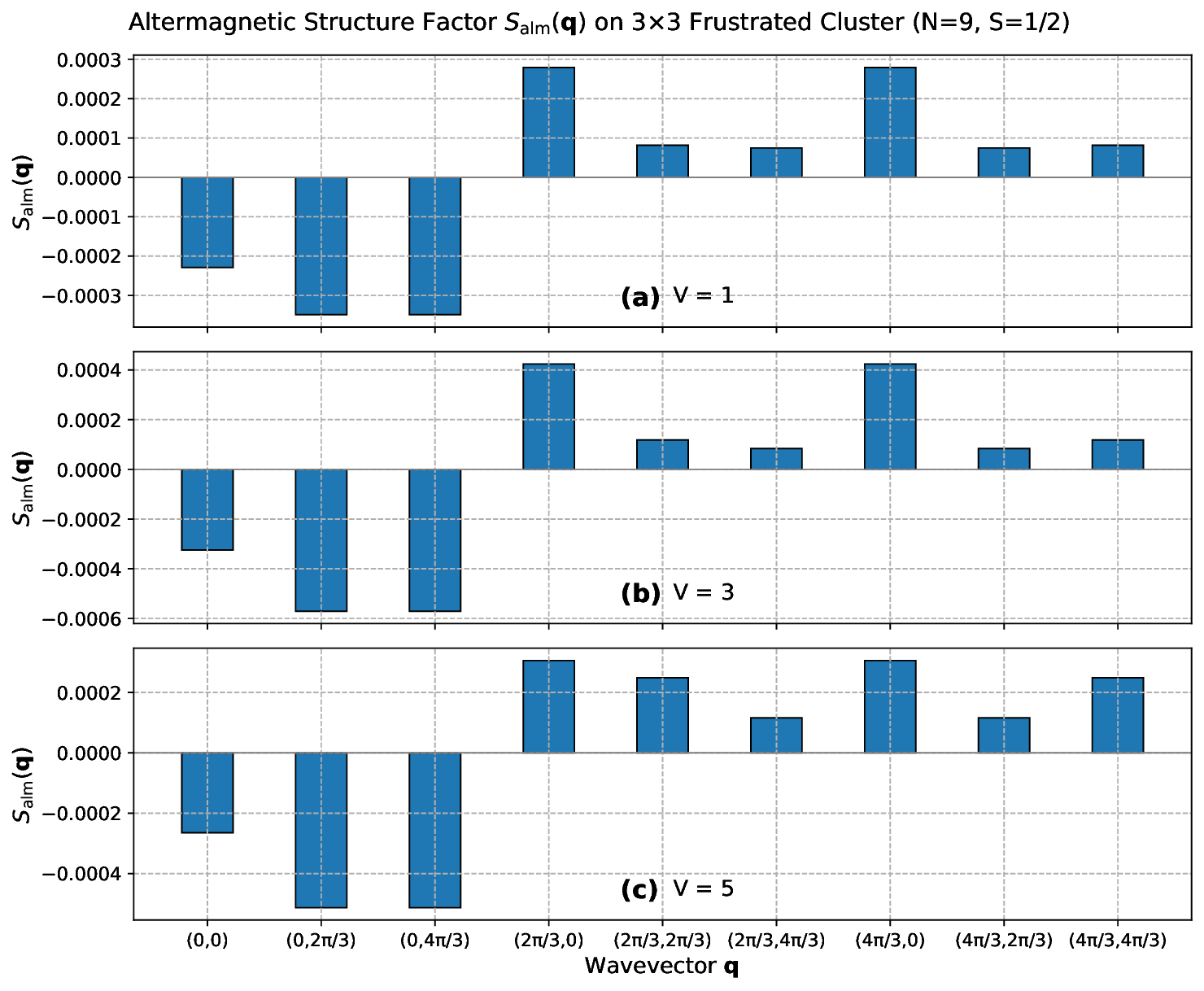}
\caption{Altermagnetic structure factor $S_{\mathrm{alm}}(\mathbf{q})$ for the half-filled ($N=9$, $S=1/2$) fully frustrated $3\times3$ lattice at $U=10$ and $V=1$, 3 and 5. The structure factor remains weak across all momenta and exhibits irregular sign variations, indicating the absence of coherent momentum-space anisotropy.}
\label{fig:Fig23}
\end{figure}
\clearpage
\begin{widetext}
\section{Ground-state degeneracy in geometrically frustrated lattice-clusters}
\label{degen}
\subsection{Simple Hubbard Model: $V=0$}
\label{degenv0}

\begin{table}[h]
\caption{Lowest five energy eigenvalues for the fully frustrated $3\times3$ lattice at half-filling ($N=9$, $S=1/2$) for different values of $U$. The near-degeneracy of the four lowest states across the interaction range indicates the presence of four-fold degenerate ground-state.}
\label{tab:degen3x3v0}
\begin{ruledtabular}
\begin{tabular}{c c c c c c}
$U$ & $E_0$ & $E_1$ & $E_2$ & $E_3$ & $E_4$ \\
\hline
0.0  & -14.99999908 & -14.99999797 & -14.99999739 & -14.99999770 & -11.99999681 \\
1.0  & -12.88387350 & -12.88387234 & -12.88387259 & -12.88387253 & -10.19536225 \\
2.0  & -10.98131341 & -10.98131298 & -10.98131373 & -10.98131396 &  -8.58928126 \\
4.0  &  -7.82410127 &  -7.82410049 &  -7.82410282 &  -7.82410100 &  -6.18109452 \\
6.0  &  -5.56230379 &  -5.56230192 &  -5.56230207 &  -5.56230205 &  -4.70841756 \\
8.0  &  -4.14421069 &  -4.14420916 &  -4.14420852 &  -4.14420757 &  -3.76437012 \\
10.0 &  -3.28579073 &  -3.28579141 &  -3.28579056 &  -3.28579242 &  -3.12106960 \\
12.0 &  -2.72702181 &  -2.72701947 &  -2.72702247 &  -2.72702111 &  -2.65853274 \\
\end{tabular}
\end{ruledtabular}
\end{table}

\begin{table}[h]
\caption{Lowest five energy eigenvalues for the partially frustrated $2\times3$ lattice at half-filling($N=6$) for different values of $U$. For $U\lesssim 3.4$ the two lowest states are degenerate within numerical precision, reflecting two‑fold degenerate ground‑state. At $U=3.6$ the degeneracy is lifted, consistent with the crossover discussed in the main text.}
\label{tab:degen2x3v0}
\begin{ruledtabular}
\begin{tabular}{c c c c c c}
$U$ & $E_0$ & $E_1$ & $E_2$ & $E_3$ & $E_4$ \\
\hline
0.0  & -7.99999996 & -7.99999997 & -7.99999998 & -6.99999996 & -6.99999994 \\
1.0  & -6.60318540 & -6.60318539 & -6.47010215 & -5.42488825 & -5.42488827 \\
2.0  & -5.41236203 & -5.41236227 & -5.25797737 & -3.99721495 & -3.99721504 \\
3.4  & -4.09388539 & -4.09388453 & -4.08888531 & -2.86184103 & -2.21357237 \\
3.6  & -3.96085239 & -3.93811357 & -3.93811358 & -2.75380606 & -2.07418654 \\
4.0  & -3.72657380 & -3.64959848 & -3.64959847 & -2.55635592 & -1.83474928 \\
6.0  & -2.86407523 & -2.59630086 & -2.59630087 & -1.85109072 & -1.16006460 \\
8.0  & -2.31082854 & -1.97924137 & -1.97924137 & -1.43414921 & -0.85726057 \\
10.0 & -1.92747500 & -1.59176818 & -1.59176817 & -1.16550924 & -0.68239588 \\
12.0 & -1.64809180 & -1.32933691 & -1.32933690 & -0.97978506 & -0.56746416 \\
\end{tabular}
\end{ruledtabular}
\end{table}
\end{widetext}

\begin{widetext}
\subsection{Extended Hubbard Model: $U=4$}
\label{degenu4v}

\begin{table}[h]
\centering
\caption{Lowest five energy eigenvalues for the partially frustrated $2\times3$ lattice at half-filling ($N=6$) for different values of $V$ at $U=4$. Near-degeneracy of the two lowest states emerge around $V\approx0.7$, coinciding with the onset of finite anisotropy in Fig. \ref{fig:Fig12}.}
\label{tab:degen2x3u4v}
\begin{ruledtabular}
\begin{tabular}{c c c c c c}
$V$ & $E_0$ & $E_1$ & $E_2$ & $E_3$ & $E_4$ \\
\hline
0.1 & -2.87585534 & -2.80712291 & -2.80712290 & -1.70537205 & -0.98985980 \\
0.4 & -0.33074430 & -0.29009049 & -0.29009050 & 0.83493894 & 1.52014536 \\
0.7 & 2.20253929 & 2.20877183 & 2.20877183 & 3.35192733 & 3.98223719 \\
0.8 & 3.03694212 & 3.03694223 & 3.04396794 & 4.18443150 & 4.78917916 \\
1.0 & 4.68478738 & 4.68478740 & 4.72162705 & 5.83730790 & 6.37737527 \\
1.6 & 9.53405474 & 9.53405461 & 9.69429318 & 10.65102699 & 10.70720592 \\
2.0 & 12.63587739 & 12.63587725 & 12.88038090 & 13.34592140 & 13.34592145 \\
\end{tabular}
\end{ruledtabular}
\end{table}

\begin{table}[h]
\centering
\caption{Lowest five energy eigenvalues for one-electron-doped ($N=10$) fully frustrated $3\times3$ lattice for different values of $V$ at $U=4$. Near-degeneracy emerges around $V\approx1.8$, consistent with the sharp increase of the anisotropy parameter in Fig. \ref{fig:Fig12}.}
\label{tab:degen3x3n10u4v}
\begin{ruledtabular}
\begin{tabular}{c c c c c c}
$V$ & $E_0$ & $E_1$ & $E_2$ & $E_3$ & $E_4$ \\
\hline
0.1 & -3.23892670 & -2.75899553 & -2.75899583 & -2.75899570 & -2.75899583 \\
0.5 & 5.24202968 & 5.68674792 & 5.68674826 & 5.68674848 & 5.68674812 \\
1.0 & 15.78627547 & 16.14030547 & 16.14030558 & 16.14030552 & 16.14030548 \\
1.7 & 30.41739380 & 30.47886363 & 30.47886402 & 30.47886417 & 30.47886392 \\
1.8 & 32.48758784 & 32.48758217 & 32.49276850 & 32.49278290 & 32.49278227 \\
2.0 & 36.46742287 & 36.46742324 & 36.46742329 & 36.46742341 & 36.58556450 \\
\end{tabular}
\end{ruledtabular}
\end{table}
\end{widetext}

\begin{widetext}
\subsection{Extended Hubbard Model: $U=10$}
\label{degenu10v}

\begin{table}[h]
\centering
\caption{Lowest five energy eigenvalues for the partially frustrated $2\times3$ lattice at half-filling ($N=6$) for different values of $V$ at $U=10$. The near-degeneracy emerging around $V\approx4.9$ coincides with the onset of finite anisotropy in Fig. \ref{fig:Fig17}.}
\label{tab:degen2x3u10v}
\begin{ruledtabular}
\begin{tabular}{c c c c c c}
$V$ & $E_0$ & $E_1$ & $E_2$ & $E_3$ & $E_4$ \\
\hline
0.1 & -1.04336672 & -0.70672146 & -0.70672146 & -0.27663232 & 0.21051055 \\
2.0 & 15.70248114 & 16.04662004 & 16.04662002 & 16.56204886 & 17.13332999 \\
4.0 & 33.18152363 & 33.44609759 & 33.44609758 & 34.08149451 & 34.62290215 \\
4.8 & 40.10709247 & 40.12915424 & 40.12915422 & 40.62449720 & 40.77117980 \\
4.9 & 40.91393515 & 40.91393516 & 40.96784765 & 41.25396653 & 41.53666263 \\
5.0 & 41.67439733 & 41.67439739 & 41.81222129 & 41.87793877 & 42.11427152 \\
\end{tabular}
\end{ruledtabular}
\end{table}

\begin{table}[h]
\centering
\caption{Lowest five energy eigenvalues for the one-electron-doped ($N=10$) fully frustrated $3\times3$ lattice for different values of $V$ at $U=10$. The emergence of near four-fold degeneracy around $V\approx4.9$ correlates with the sharp increase in the anisotropy parameter shown in Fig. \ref{fig:Fig17}.}
\label{tab:degen3x3n10u10v}
\begin{ruledtabular}
\begin{tabular}{c c c c c c}
$V$ & $E_0$ & $E_1$ & $E_2$ & $E_3$ & $E_4$ \\
\hline
0.1 & 5.49492154 & 6.11242872 & 6.11242882 & 6.11242899 & 6.11242894 \\
2.0 & 46.62199045 & 47.27474681 & 47.27474674 & 47.27474698 & 47.27474695 \\
4.0 & 89.38003343 & 89.87925491 & 89.87925475 & 89.87925541 & 89.87925518 \\
4.8 & 106.15884558 & 106.22807136 & 106.22807125 & 106.22807134 & 106.22807126 \\
4.9 & 108.20932540 & 108.20932555 & 108.20932474 & 108.23487481 & 108.23487479 \\
5.0 & 110.17346578 & 110.17346534 & 110.17346552 & 110.17346524 & 110.30525841 \\
\end{tabular}
\end{ruledtabular}
\end{table}
\end{widetext}
\bibliography{manuscript}
\end{document}